\documentclass[letterpaper,11pt]{article}

\usepackage{geometry}
\usepackage{graphicx}
\usepackage{aas_macros}
\usepackage[super]{natbib} % bibliography
\usepackage{comment}
\usepackage{url}
\usepackage{hyperref}
\usepackage{fancyhdr}
\usepackage{pdfpages}

\usepackage{sidecap}   
\usepackage{wrapfig}
\usepackage{amsmath,amssymb}
\sidecaptionvpos{figure}{t}

\def\Hitomi{{\em Hitomi}}
\def\hitomi{{\em Hitomi}}

\def\xmm{{\em XMM-Newton}}

\begin{document}

\includepdf[]{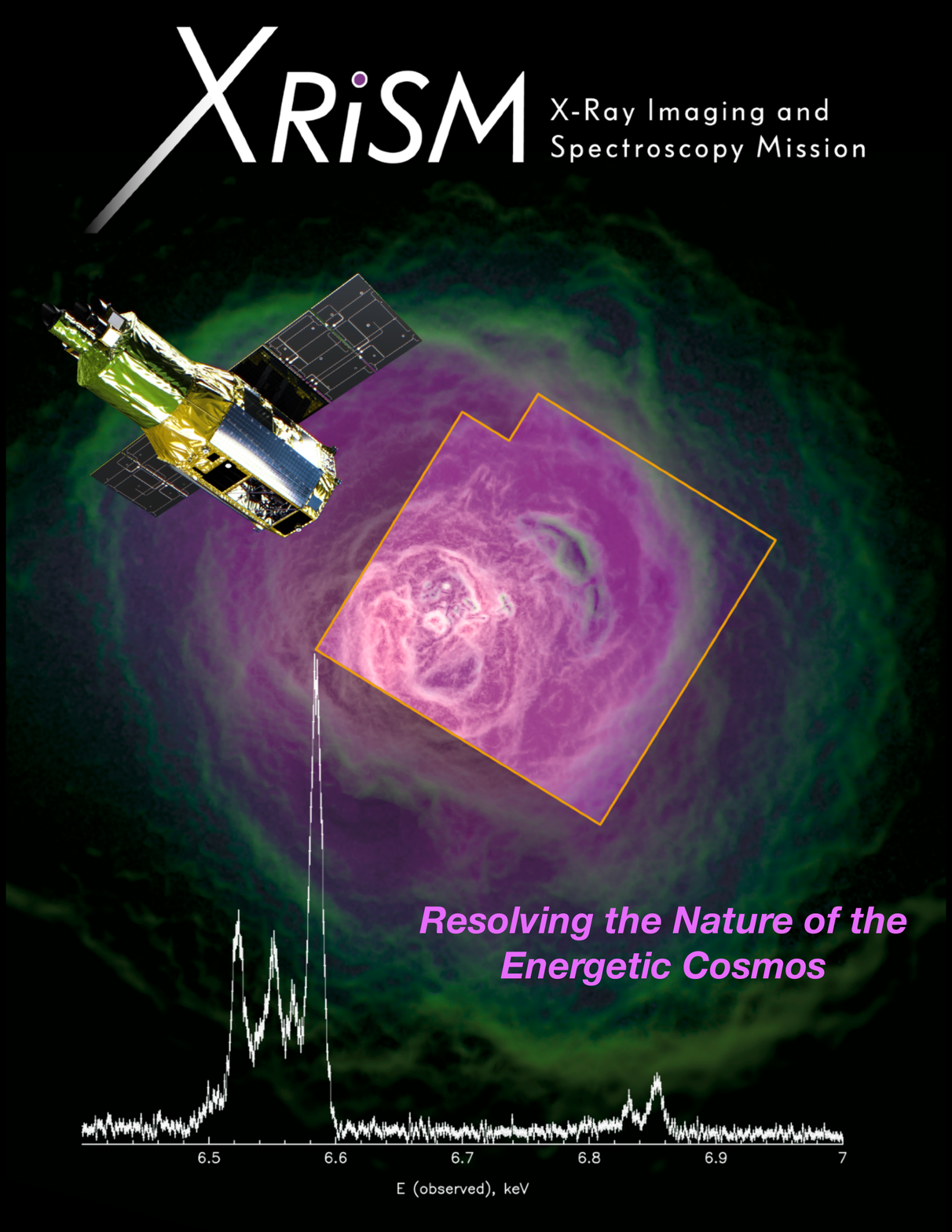}
\includepdf[]{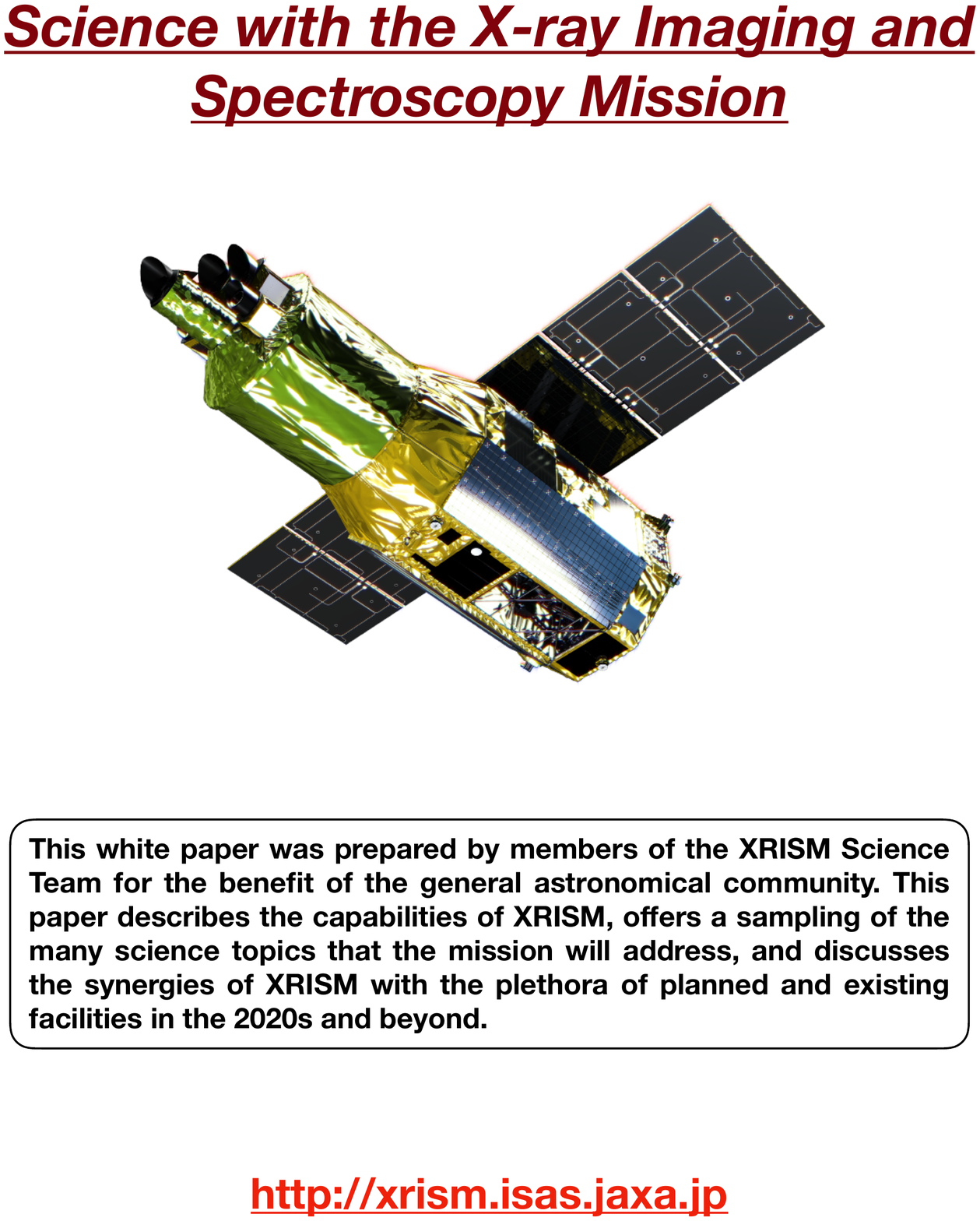}

\pagenumbering{roman}

\renewcommand{\contentsname}{}
\tableofcontents

\clearpage

\setcounter{page}{1}
\pagenumbering{arabic}

\section{Introduction}

The {\it X-ray Imaging and Spectroscopy Mission (XRISM)}, formerly known as the {\it X-ray Astronomy Recovery Mission (XARM)}, is a JAXA-NASA collaborative mission with ESA participation. {\it XRISM} will offer non-dispersive, high-resolution X-ray spectroscopy in the soft X-ray bandpass ($\sim 0.3-12$ keV), while offering complementary CCD imaging resolution over a wide field of view. The primary purpose of the mission is to recover the science that was lost after the {\it Astro-H/Hitomi} mission failed in 2016, approximately one month after launch. Four science categories have been defined for {\it XRISM}: (1) Structure formation of the Universe and evolution of clusters of galaxies; (2) Circulation history of baryonic matter in the Universe; (3) Transport and circulation of energy in the Universe; (4) New science with unprecedented high-resolution X-ray spectroscopy. A full description of the mission is provided in an SPIE proceeding\cite{tashiro18}; the reader is referred there for more detail.

\begin{figure*}[htb]
\begin{center}
\includegraphics[width=0.9\textwidth]{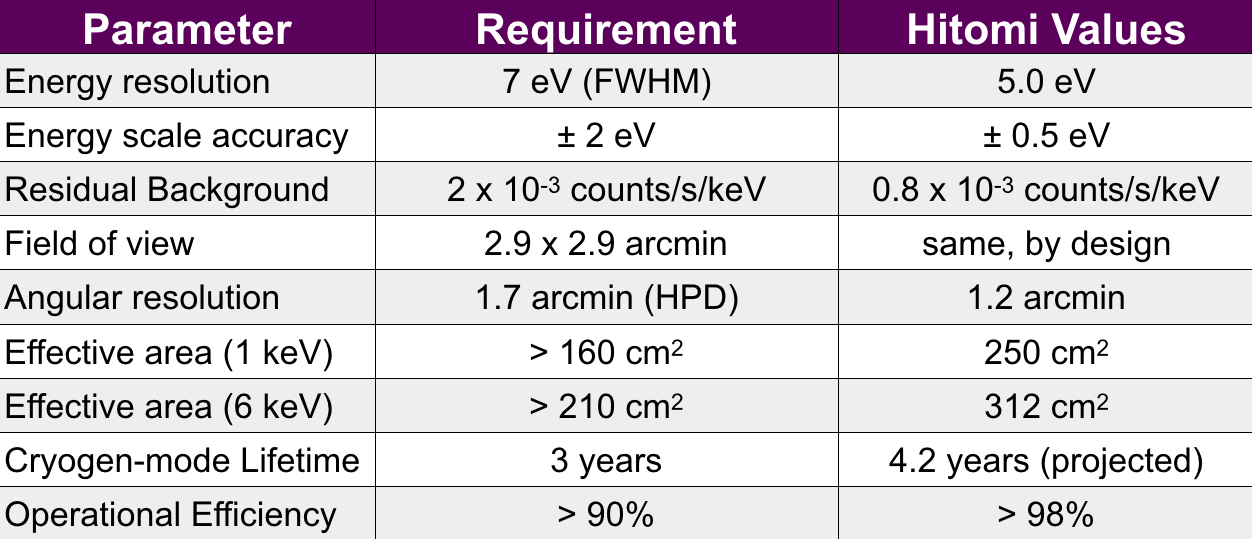}
\end{center}
%\vspace{-0.2in}
\caption[]{\footnotesize{The mission requirements for both the {\it XRISM} Resolve instrument and the {\it Hitomi} Soft X-ray Spectrometer (SXS), along with the on-orbit performance of the {\it Hitomi} mission as measured in 2016. {\it Hitomi} met or exceeded requirements in all areas.}}
\label{fig:table}
\vspace{-3mm}
\end{figure*}

{\it XRISM} represents a revolutionary leap forward in X-ray spectroscopy. With a spectral resolution 20-40 times better than the CCD instruments that are used on {\it Chandra, XMM-Newton}, and {\it Suzaku}, as well as a substantially increased collecting area and bandpass over the grating instruments on those missions, {\it XRISM} represents a new tool for studying the universe, and will enable science never before possible. {\it XRISM} will do this using two instruments: 

\begin{itemize}
    \item {\it Resolve}, a soft X-ray spectrometer with a constant $<7$ eV FWHM spectral resolution over the entire bandpass. {\it Resolve} has a field of view of $2.9' \times 2.9'$ over an array consisting of a $6 \times 6$ pixel X-ray microcalorimeter (pixels are $30"$ in size). The array has an operating temperature of 50 mK, and must be cooled by a multi-stage adiabatic demagnetization refrigerator. This system can operate in both cryogen and cryogen-free modes, meaning that the mission can continue to operate after the exhaustion of the liquid helium cryogen supply. The mission requirements for {\it Resolve} are given in Figure~\ref{fig:table}, along with the measured values from the in-flight performance of {\it Hitomi} (all of which exceeded requirements).
    
    \item {\it Xtend}, a soft X-ray imager providing simultaneous coverage of the {\it Resolve} field and the surroundings over a $38' \times 38'$ field of view. {\it Xtend} will provide CCD-quality imaging spectroscopy similar to that available on {\it Suzaku} (energy resolution $<250$ eV at 6 keV), with $\sim 1'$ spatial resolution.
\end{itemize}

\begin{figure*}[htb]
\begin{center}
\includegraphics[width=1.0\textwidth]{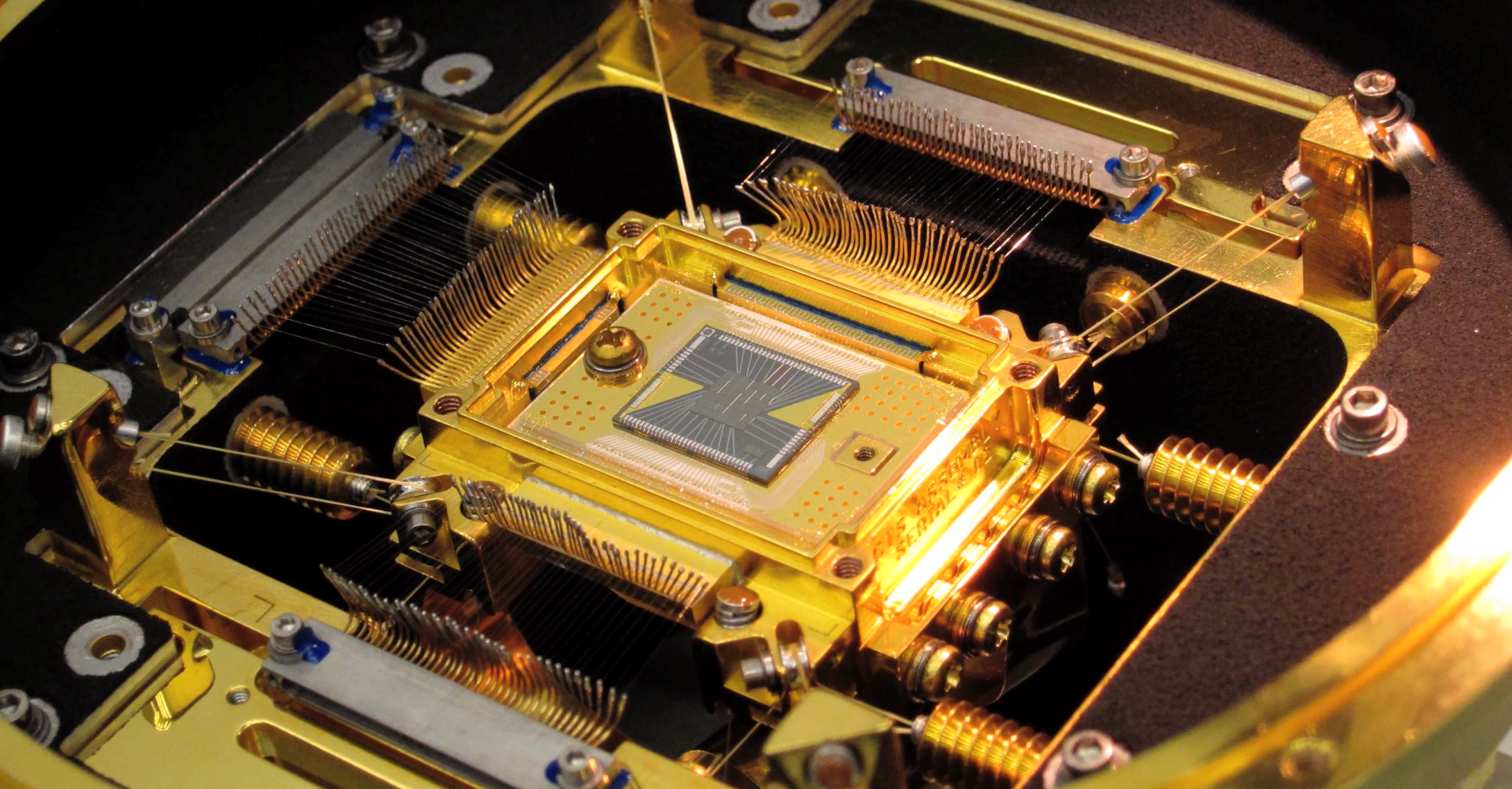}
\end{center}
%\vspace{-0.2in}
\caption[]{\footnotesize{The {\it Hitomi} Soft X-ray Spectrometer, photographed prior to launch of that mission in 2016. {\it XRISM's Resolve} instrument will be identical.}}
\label{resolve}
\vspace{-2mm}
\end{figure*}

In Figure~\ref{resolve}, we show the {\it Hitomi} SXS instrument, mounted in the calorimeter spectrometer insert. The {\it XRISM} instrument is currently in production, but will be identical to this. The pixels in the detector measure the heat imparted by incoming X-ray photons (only a fraction of a degree, hence the need for the detector to be cooled to such a low temperature). From this temperature change, the energy of the incoming X-ray photon can be calculated.

\begin{figure*}[htb]
%vspace{-0.2in}
\begin{center}
\includegraphics[width=1\textwidth]{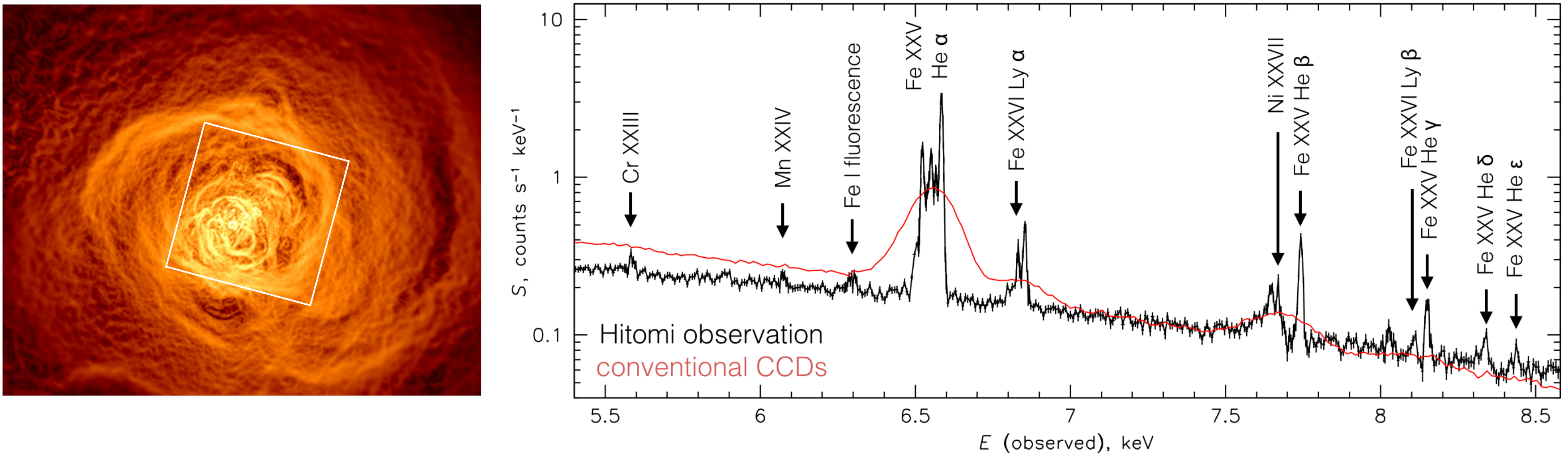}
\end{center}
\vspace{-0.2in}
\caption[]{\footnotesize {{\bf Left:} {\it Chandra} X-ray image of the Perseus cluster core, filtered to emphasize structures in the hot gas \cite{Wal17}. {\bf Right:} spectrum from the Perseus core (white square in the left panel) observed with {\it Hitomi's} microcalorimeter (black) and {\it Suzaku's} CCD imaging spectrometer (red). {\it XRISM} will provide similar high-resolution spectra in the $0.3-12$ keV band for extended X-ray sources \cite{Hit16}.}}
\label{fig:hit_suz}
\vspace{-2mm}
\end{figure*}

{\it XRISM} is particularly well-suited to the astrophysical analysis of extended objects, such as galaxy clusters and supernova remnants. Because the light is not dispersed, spatial information about these sources is not lost as it often is in observations with the gratings on {\it Chandra} or {\it XMM-Newton}. {\it Resolve} features an exceptionally low background of $\sim 1$ count day$^{-1}$ 10 eV$^{-1}$. As an example of this, in the {\it Hitomi} spectrum of the Perseus Cluster shown in Figure~\ref{fig:hit_suz}, there are $\sim 21,000$ counts in the Fe K$\alpha$ complex shown; of these, only 16 counts are due to the background.

{\it XRISM} is planned for launch in 2022 from Tanegashima Space Center, aboard a JAXA HII-A rocket to an orbital altitude of 550 km and an inclination of 31 degrees. Two identical X-ray Mirror Assemblies (one for {\it Resolve} and one for {\it Xtend}), each with 203 nested circular foils in both the primary and secondary mirrors, focus X-rays onto the focal plane detectors at the base of the spacecraft, 5.6 meters away. The mirrors have an angular resolution of order $1'$.

In this white paper, we give a brief introduction to the science that will be enabled by {\it XRISM}. This work is by no means exhaustive. A more full description of the scientific topics presented here can be found in a collection of 17 white papers written in 2014, prior to the launch of {\it Hitomi} \footnote{\href{https://heasarc.gsfc.nasa.gov/docs/hitomi/about/paper_list.html}{https://heasarc.gsfc.nasa.gov/docs/hitomi/about/paper\_list.html}}. This white paper serves as a summary of those more detailed white papers, with the important inclusion of science learned from {\it Hitomi} and a more timely description of {\it XRISM} in the landscape of astronomy in the 2020s.

\section{GALAXY AND CLUSTER EVOLUTION}
{\it (Primary editors: Irina Zhuravleva, Erin Kara, Chris Done)}\\

One of the biggest questions in modern astrophysics and cosmology is how the largest structures in the Universe form and evolve with time. As a result of enormous efforts in observational and theoretical astrophysics, we now understand what processes are playing an important role in the evolution of galaxies, groups, and clusters. However, the physics of the involved processes and the dynamical and chemical structure of the largest objects and their environments are still not well understood. 

Most of the visible matter in the Universe is in the form of hot, X-ray emitting gas within massive galaxies, groups, and clusters of galaxies as well as in the interstellar and intergalactic gas between them. X-ray observations have provided spectacular high-resolution images of these objects. These data revealed a wealth of structures associated with various complex physical processes that take place in these environments. With {\it XRISM}, we will be able to resolve individual spectral lines and trace gas motions through the Doppler broadening and line energy shifts in many galaxies and clusters of galaxies. We will robustly measure the chemical composition, multi-temperature structure, and ionization state of the hot gas. 

%The major questions in galaxy and cluster evolution that {\it XRISM} will address include:
%\begin{itemize}
%	\item How do galaxies and clusters of galaxies sustain themselves against gravity? What processes are shaping the evolution of large-scale structures beyond pure gravitational collapse? How is the gravitational energy released and converted into heat?
%	\item What is powering supermassive black holes (SMBHs)? How do SMBHs exchange energy with galaxies and clusters, and how does it affect the evolution of galaxies on small and large scales?
%	\item What is the chemical make up of matter in the Universe and how does it change with time? What processes are enriching the gas at different stages of evolution?
%\end{itemize}

In this chapter, we briefly review the key {\it XRISM} science related to the evolution of galaxies and clusters, physics of active galactic nuclei (AGN) and AGN feedback as well as the chemical composition and evolution of large-scale structure. A detailed description and simulations of astrophysical sources can be found in the Astro-H white paper \citep{Kit14}.

\subsection{Dynamics of the most energetic events in the Universe}
Clusters of galaxies - the largest and most massive structures in the Universe - are formed hierarchically via a sequence of mergers and accretion of surrounding matter, including smaller clusters and galaxy groups. These processes drive shocks in the intracluster medium (ICM), trigger bulk and turbulent motions of the gas, and channel energy into magnetic fields and relativistic particles \citep{Kra12,Byk15}. On smaller scales, an interaction between the energy generated by the accretion of matter onto a supermassive black hole (SMBH) and gas in the host galaxy (AGN feedback) takes place. Even a small fraction of the energy produced in the vicinity of a black hole and transferred to the gas has a profound effect on the host galaxy and cluster evolution \citep{Fab12,Byk15,Wer19}. {\it XRISM} will probe the kinematics of such complex and energetic processes for the first time, transforming our ``static," single-frame view of these processes  into a full dynamic picture.

{\bf Dynamics of violent mergers.} Mergers of galaxy clusters are the most energetic events in the Universe since the Big Bang, with the total kinetic energy of colliding clusters reaching $10^{65}$ ergs \citep{Mar99}. The hot intracluster gas that fills the gravitational potential wells of clusters is strongly affected by these collisions and can be traced with X-ray observations. 

Strong shocks, bulk motions of the gas, and turbulence dissipate energy from collisions, heating the gas to very high temperatures. Shocks have been robustly detected in a handful of clusters using high-resolution X-ray images  \citep{Mar07,Rus10,Aka13,Ich15,Uch16,Can17,Ued18}. However, the dominant energy conversion mechanism  - via bulk motions of the gas and turbulence - has never been measured. Resolve will provide such measurements in many mergers at different stages thanks to its superb spectral resolution. Combining this velocity information with other existing data (X-ray, optical, sub-mm, radio) and numerical simulations, we will be able to reconstruct the geometries and dynamical stages of merging systems. Furthermore, the large field of view of Xtend (a factor of $\sim$ 2.2 larger than {\it Chandra}/ACIS-I) will provide complementary measurements of radial profiles of X-ray surface brightness, gas density, and temperature up to large radii.

Mergers are often accompanied by ``cold fronts" (e.g., A2319, A3667, A754) that appear as sharp contact discontinuities in the X-ray surface brightness. Cold fronts arise either from subcluster stripping by ram pressure or from ``sloshing" of the low-entropy gas in a disturbed gravitational potential well \citep{Markevitch00,Vik01,Mar07,Roe11,ZuH11,Nul13,Wer16,ZuH16}. In the case of gas stripping, bulk motions of the subclusters are mostly in the plane of the sky, while in the sloshing case, there is a near-sonic flow along the line of sight inside the front. Resolve will measure both bulk and turbulent motions of the gas in such cold fronts, providing a solid basis for probing the microphysics of the ICM using cold fronts.

Many merging clusters host giant diffuse radio halos filled with relativistic particles. These particles emit synchrotron radiation in the cluster magnetic field \citep{Fer12}, rapidly losing their energy and mixing with the thermal plasma on length-scales well below the sizes of radio halos. Therefore, in order to explain the observed extent of these halos, particles have to be constantly accelerated in situ. One way to accelerate particles is by scattering on magnetosonic waves produced by turbulence \cite{Bru14}. Resolve observations will test the  theories of cosmic-ray acceleration by tracing turbulence in regions within clusters that have strong radio emission. 

{\bf Precise mass measurements and cluster cosmology.}
Knowing masses of galaxy clusters  at different redshifts, we can trace the growth of structures over time and constrain cosmological models. This is usually done through measurements of cluster mass function, which decreases exponentially at the high-mass end. This exponential ``tail" is very sensitive to the dark energy density parameter and provides a way to infer the role of dark energy in cosmic expansion \citep{All04,Vik09}.       

\begin{figure*}[htb]
\begin{center}
\includegraphics[width=1\textwidth]{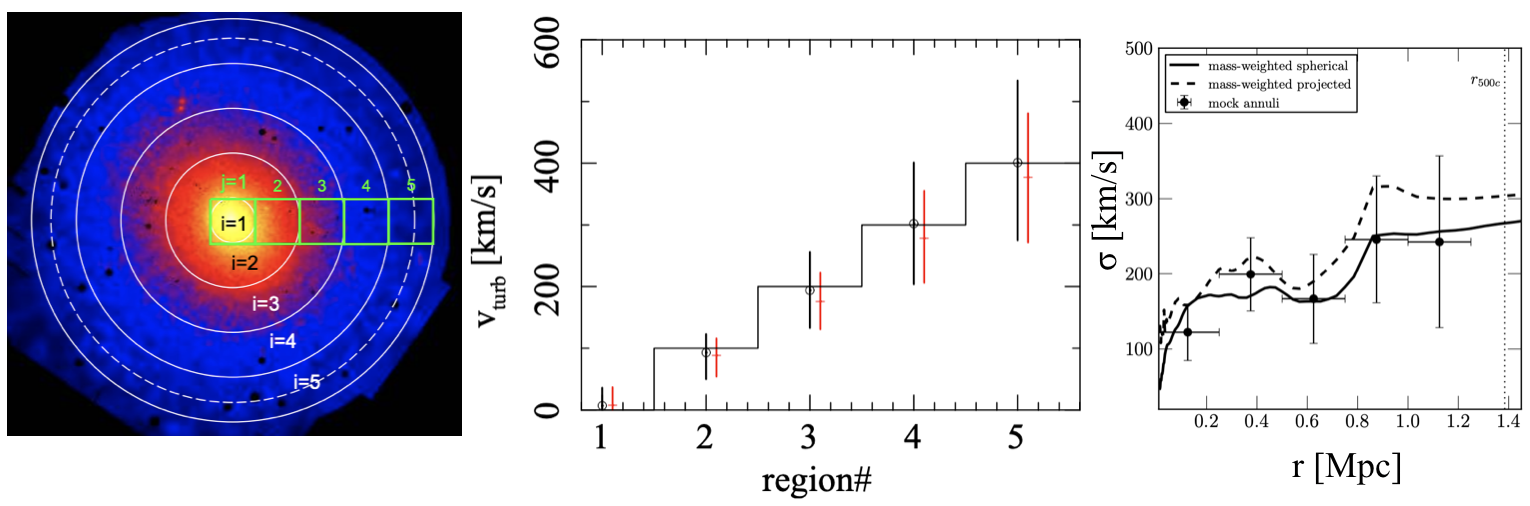}
\end{center}
\vspace{-0.2in}
\caption[]{\footnotesize {An illustration of {\it XRISM} reconstructing radial profiles of the velocity of gas motions. {\bf Left:} {\it XMM-Newton} image of A2199 and $3'\times3'$ {\it XRISM} pointing regions (green boxes) up to $r_{2500}$ (dashed circle). {\bf Middle:} the input (solid line) and recovered (points) velocity dispersion in A2199. Black points: simultaneous fitting of the spectra from all annuli; red crosses: separately fitting the spectrum in each annulus. Errors are at 90\% confidence. Adapted from \cite{Kit14}. {\bf Right:} Measurements of gas velocity dispersion as a function of radius based on the mock data of a relaxed galaxy cluster from cosmological simulations. Curves: true velocity dispersion from simulations; points: recovered dispersion from the mock Resolve spectra. Adapted from \cite{Nag13}.}}
\vspace{-2mm}
\label{fig:mbias}
\end{figure*}

The precision of cluster cosmology depends on the accuracy of the mass measurements. One of the most important ways of measuring cluster masses is based on the assumption that the thermal pressure alone supports the hot gas against the cluster's gravity (hydrostatic equilibrium). However, numerical simulations show that the macroscopic motions of the gas may provide significant non-thermal pressure support, up to 25-30 \% at r$_{200}$ \citep{Ras06,Lau09,Zhu13,Nel14,Shi16,Vaz18}. {\it Hitomi} observations of the Perseus core, which is dominated by the central massive galaxy NGC~1275, showed that the kinetic pressure is less than 10\% of the thermal pressure on scaless of 60 kpc \citep{Hit16}. {\it XRISM} will continue {\it Hitomi's} legacy by measuring velocities of gas motions and the non-thermal pressure contribution with an accuracy of a few percent in many relaxed, bright galaxy clusters (Fig. \ref{fig:mbias}). Velocity measurements up to $\sim$r$_{2500}$, for even a small sample of nearby relaxed clusters, will significantly improve the constraints on dark energy and other cosmological parameters.

{\bf Physics of cool cores and mechanical AGN feedback.} SMBHs residing in galaxy clusters \citep{EHT19} are preventing strong cooling and vigorous star formation in their host galaxies through feedback processes \citep{Chu00,McN00,Fab12}. These processes include the inflation and rise of bubbles of relativistic particles, generation of shocks and sound waves, uplift of cooler, multi-phase gas, and turbulence and other motions of the gas \citep{Boe93,Chu01,Fab03,San04,For07,McN07,Sim08,Wer10,Fab11, Are16,Zhur18}.  

\begin{figure*}[htb]
\begin{center}
\includegraphics[width=1\textwidth]{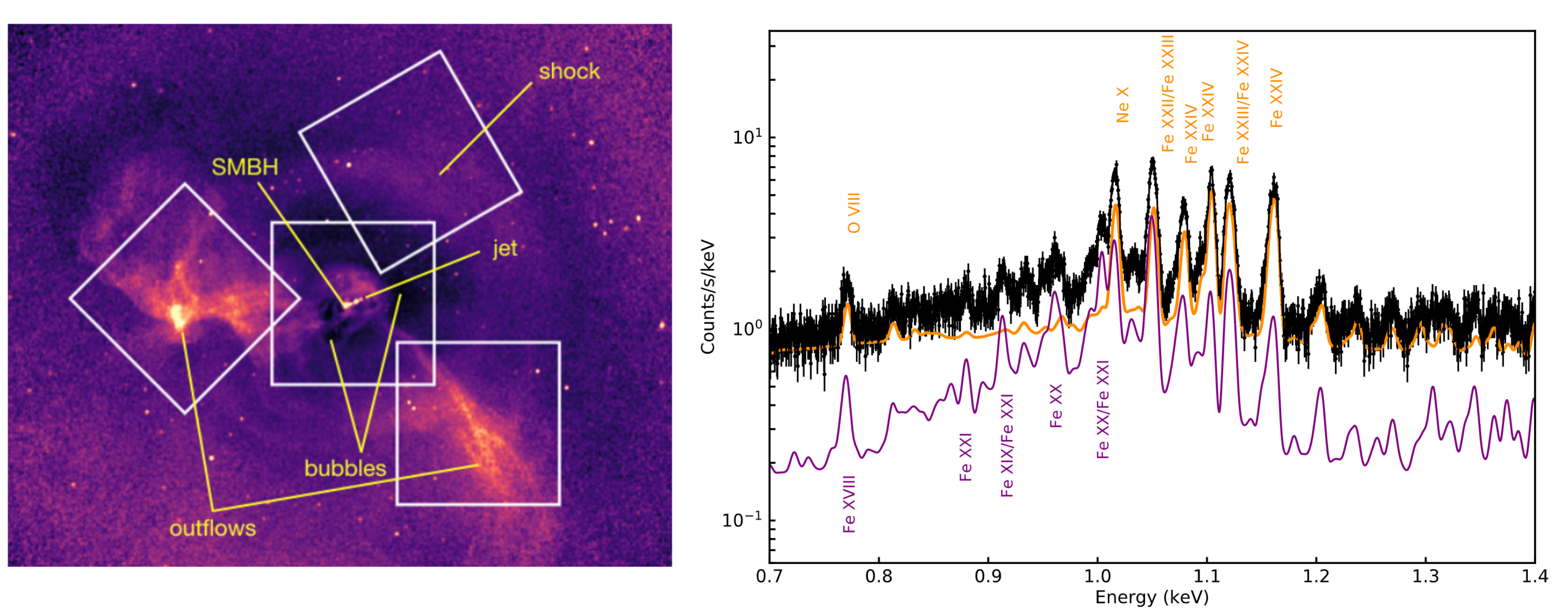}
\end{center}
\vspace{-0.2in}
\caption[]{\footnotesize {Spatially and spectrally resolving AGN feedback with {\it XRISM}. {\bf Left:} {\it Chandra} X-ray image of the core region of the Virgo cluster (dominated by the central galaxy M87) divided by a spherically-symmetric model of surface brightness. The image highlights the complexity of structures in the gas produced by the AGN feedback \citep{For07,For17}. Each white box ($3'\times 3'$) shows the field of view of Resolve. Two regions are pointed at the cool and bright structures, one at the shock and another one at the center of M87. {\bf Right:} a 100 ks simulated Resolve spectrum of one of the regions with outflows (left box) in M87. The spectrum shows the contributions from the different thermal components to the Fe L line complex. Adapted and updated from \cite{Kit14}.}}
\label{fig:m87}
\vspace{-2mm}
\end{figure*}

%%Hitomi observations of the core of the Perseus cluster revealed that velocity dispersion is the largest ($\sim 200$ km s$^{-1}$) toward the central AGN and the AGN-inflated bubble in the NW. Elsewhere, the velocity dispersion is $\sim 100$ km s$^{-1}$, and the line-of-sight velocity gradient is also $\sim 100$ km s$^{-1}$ \citep{Hit18b}. Interestingly, such velocities are compatible with those predicted in cosmological numerical simulations \citep{Bou17,Lau17}, simulations of gas sloshing \citep{ZuH18}, and AGN feedback \citep{Hil17,Lau17}. Hitomi results are also consistent with both the shock-heating \citep{Fab17,Hit18b} and turbulent heating scenarios \citep{Hit18b,Zhur18} of cluster cool cores. 

Resolve will perform comprehensive velocity diagnostics in the regions dominated by AGN feedback physics in the brightest cool-core clusters. Such measurements will address the role of turbulence in cooling-heating balance and metal mixing. In nearby objects like the Virgo cluster (Figure~\ref{fig:m87}), Resolve will measure the kinetic energy associated with AGN-injected bubbles and uplifted gas filaments.

\subsection{The Physics of Black Holes and Active Galactic Nuclei}
{\it XRISM} will revolutionize how we understand accreting SMBHs, starting from how cold gas flows in from the interstellar medium (ISM) of the galaxy, down to the inner accretion flow where matter is affected by the strong gravity of the black hole. This gravitational inflow leads to massive outflows, which may be powerful enough to prevent gas from cooling to form stars, thus affecting the evolution of the host galaxy. X-ray continuum emission is produced within a few gravitational radii of the central black hole and irradiates the remaining infalling or outflowing gas (e.g., the accretion disk, broad-line region clouds or the dusty torus), producing a plethora of emission features, known colloquially as the reflection spectrum. The Fe K$\alpha$ line is the most commonly used diagnostic of the structure around the AGN because iron is an abundant element and this line grouping appears in a relatively isolated region of the spectrum. {\it XRISM} will provide an unprecedented view of this important spectral feature. The combination of improved sensitivity and better spectral resolution will enable XRISM results to be extended to large samples of objects.

\begin{figure}[h]
\centering
\includegraphics[width=\textwidth]{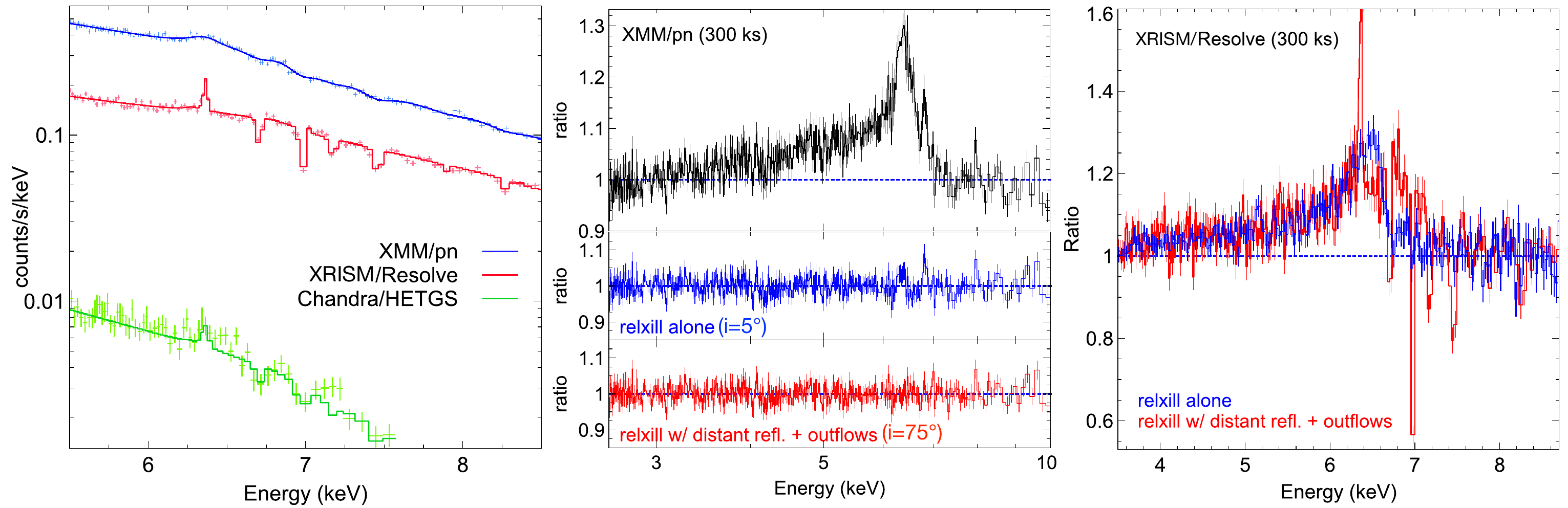}
\caption[]{\footnotesize {{\bf Left:} Fe K$\alpha$ line profile of MCG$-$6-30-15 as seen in 300~ks by {\it XMM-Newton} pn (blue) and {\it Chandra} HETGS (green), along with a Resolve simulation in red. {\bf Middle:} The ratio of the {\it XMM} pn spectrum to a power-law model fit between 2--4~keV and 7.5--10~keV. The pn spectrum can be equally well described by pure relativistic reflection with an inclination of $5\deg$ (as shown in blue) or by relativistic reflection with an inclination of $75\deg$ with additional emission and absorption, including an ultrafast outflow of $v_{out} \sim 0.13$c (as shown in red). {\bf Right:} {\it XRISM} will easily distinguish between these two scenarios, as the narrow emission and absorption will be resolved.}}
\label{fig:bhspin}
\vspace{-2mm}
\end{figure}

\textbf{Black hole spin through observations of broad, ionized reflection:} Japan's {\it ASCA} satellite opened a new field in X-ray astronomy with the discovery of the broad Fe~K$\alpha$ line in MCG$-$6-30-15 \citep{tanaka95}. The implication of the broad line is immense, as it may provide a way of measuring an elusive fundamental parameter of a black hole: its spin \citep{fabian89}. As such, the broad Fe K$\alpha$ line has ignited healthy debate and continued interest \cite{hagino16}. Rapid time variability, X-ray reverberation mapping \cite{zoghbi12,kara16}, and quasar microlensing \cite{chartas17} indicate that the broad Fe K$\alpha$ line is produced by reprocessing off the inner accretion disk within $\sim10 R_g$, but the details of the inner disk radius and kinematics of the flow are still highly debated. By disentangling contributions from other spectral features (including narrow emission from the outer disk and/or torus, or absorption though ionized outflows along our line of sight), we can confidently probe properties of the inner accretion flow to make a clean estimate of the black hole spin. {\it XRISM} will give us a clear understanding of how much material is actually making it to the black hole and how much material is being removed through massive outflows (see the case study of MCG$-$6-30-15 in Figure~\ref{fig:bhspin}). Furthermore, the combined use of Resolve and Xtend will allow for complementary measurements of the short light-travel time delays associated with the relativistically broadened Fe~K$\alpha$ emission line (known as X-ray reverberation).  

\textbf{Outer disk/torus structure through observations of narrow, neutral reflection:} A narrow Fe line at 6.4~keV is nearly ubiquitous in AGN spectra. It is due to neutral Fe~K$\alpha$ fluorescence off circumnuclear optically-thick material, but its physical association with, for example, the dusty torus that may serve as a gas reservoir to the disk, or the outer edge of the accretion disk, is still not fully understood. {\it Hitomi} observed 1 AGN, the low-luminosity AGN NGC~1275 at the center of the Perseus Cluster, and found a narrow Fe~K$\alpha$ line, with a 500-1600~km s$^{-1}$ width that suggests it is associated with a molecular torus extending tens to hundreds of parsecs \citep{hitomi18a}. {\it XRISM} will extend these studies to a large sample of AGN, spanning a range of mass accretion rates. This is important, as Seyfert galaxies observed with {\it Chandra} High-Energy Transmission Grating (HETG) suggest higher-velocity structure in the Fe~K$\alpha$ line, placing the Fe~K emitting region within the optical broad line region \citep{gandhi15,minezaki15}. Moreover, recent advances using 3rd-order spectra from {\it Chandra} HETG find that the narrow Fe~K$\alpha$ line in one Seyfert 1 galaxy NGC~4151 is, in fact, asymmetrically broadened, suggestive of Doppler shifts and weak gravitational effects \citep{miller18}. This allows for determination of both the radius and inclination of the narrow Fe~K$\alpha$ emitting region. In this one tantalizing case, the outer and inner disk inclinations do not match, which might suggest a change in the vertical scale height of the disk at large radii due to radiation pressure. Interestingly, similar interpretations of changes in vertical disk structure have also been suggested \citep{gardner17} to explain intensive disk reverberation mapping of NGC~4151 \citep{edelson17}. 

\subsection{Chemical Composition and Evolution in the Universe}

The chemical make-up of the universe represents the sum of billions of past supernovae. In contrast to stellar systems, H~II regions, or planetary nebulae, gas in the ICM is largely free of the effects of optical depth, extinction, and non-equilibrium ionization. In such a physical environment, robust measurements of chemical enrichment can be done with the high-resolution spectra from Resolve, providing valuable insights into the production of heavy elements by early stellar populations.

\begin{figure}[htb]
\centering
\includegraphics[width=1.0\textwidth]{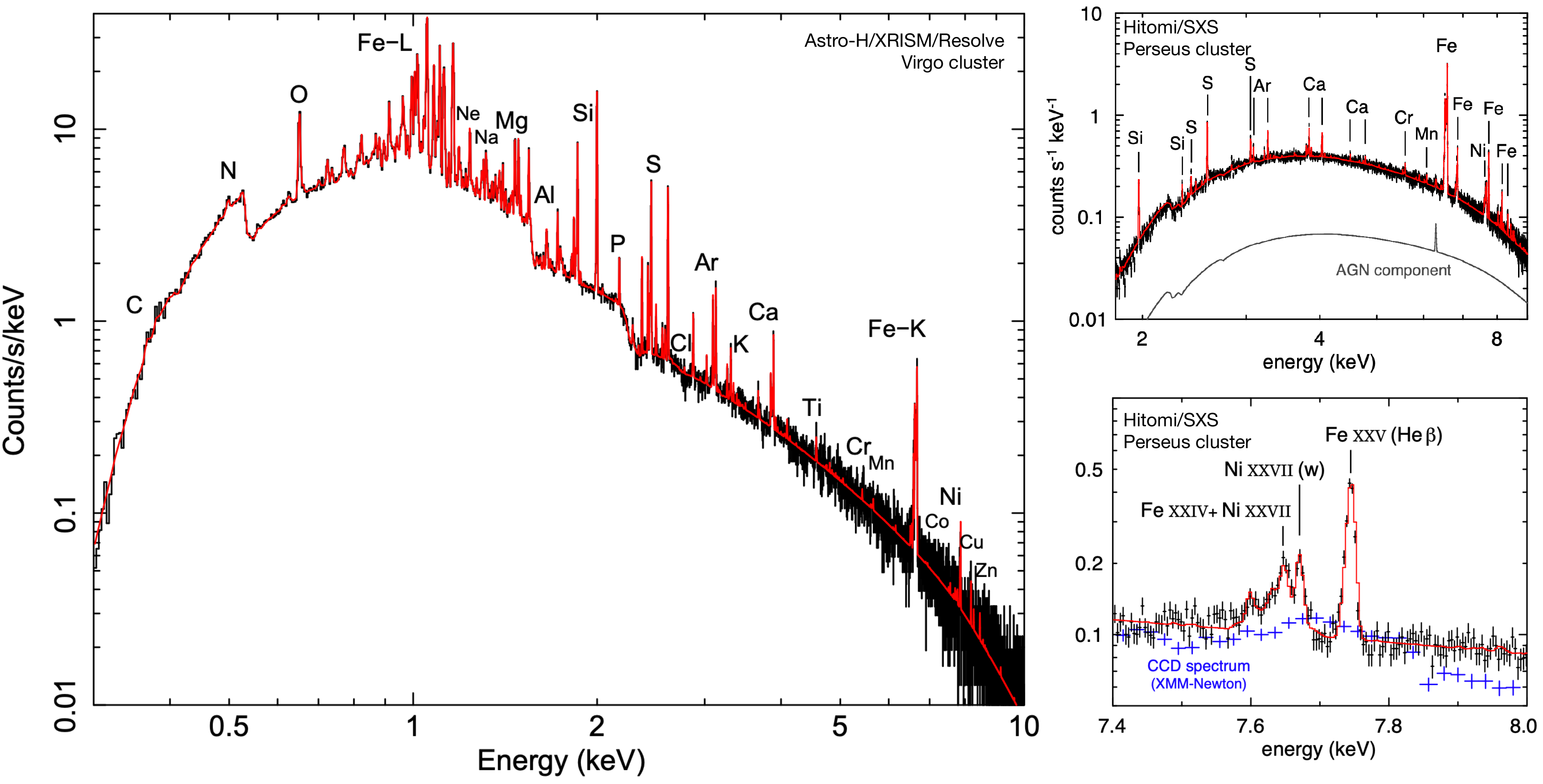}
\caption{\footnotesize {{\bf Left:} {\it XRISM} simulated spectrum (black) of the core of M87 as observed with 100 ks and the best-fitting model (red). Adapted from \cite{Kit14} {\bf Right:} {\it Hitomi} spectrum from the core of the Perseus cluster. Broad-band spectrum is shown on the top panel (note that the gate valve was in place during the observation and it absorbs most X-rays below $\sim 2$ keV), while the zoomed-in on the high-energy end spectrum is shown in the bottom panel. The best-fitting model is shown in red. {\it Hitomi} resolved individual lines in the spectrum (black vs. blue points)\cite{Hit17}. }}
\label{fig:chem}
\vspace{-2mm}
\end{figure}

{\bf Chemical composition of bright nearby galaxies and galaxy clusters.} All elements between C and Zn show emission lines between 0.1 and 10 keV in spectra of galaxy clusters. Resolve will measure the abundances of the $\alpha$-elements and Fe with remarkable accuracy of a few percent in nearby cool-core clusters. These measurements alone already provide tight constraints on the mechanisms that produce these chemical elements \citep{Mat03,deP07,Sim09, Mer18}. With sufficiently long exposures, Resolve will detect weak lines from rare elements in the ICM, such as Na, Al, Cr, and Mn (Figure~\ref{fig:chem}). Al and Na are sensitive tracers of the metallicity of the underlying stellar population, while Cr and Mn can be used to probe the characteristics (including metallicity) of SNIa progenitors, resulting in a wealth of information about the chemical enrichment history of the ICM and the Universe. 

Outside the regions dominated by the central brightest cluster galaxy, metallicity measurements will be important for constraining the chemical evolution driven by the remaining cluster member galaxies, which may have witnessed a different star formation history over cosmic time. By measuring metal abundances in galaxies and groups of galaxies, {\it XRISM} will further investigate the mass dependence of the chemical enrichment history \citep{Kit14}.

{\it Hitomi} observations of the Perseus core confirmed the power of high-resolution spectroscopy: the Cr/Fe, Mn/Fe and Ni/Fe ratios were measured accurately for the first time (Figure~\ref{fig:chem}). This measurement revealed that these elements have a near-solar abundance ratio. Comparison with the nucleosynthesis calculations disfavoured the scenario of the SNIa progenitors being exclusively white dwarfs with masses below the Chandrasekhar limit \citep{Hit17}. Combining {\it Hitomi} measurements with {\it XMM-Newton} RGS grating spectra at lower energies yielded reliable measurements of the abundances of O, Ne, Mg, Si, S, Ar, Ca, Cr, Mn, Fe, and Ni, which were used to further test existing SN nucleosynthesis calculations. These results showed, in particular, that core-collapse supernova (CCSN) yield calculations with included neutrino physics improve the agreement with the observed pattern of $\alpha$-elements \citep{Sim19}. 

{\bf Chemical evolution in high-redshift Universe.} Tracing the star formation at different redshifts is directly connected to the chemical evolution of the Universe, so measuring abundances of different elements at high redshifts is crucial. Based on models of galaxy formation, the relative population of massive stars, and consequently the rate of CCSN, is thought to increase in the high-redshift Universe. The ratio of abundances of Fe and light alpha-elements like O, Mg, and Si can check this prediction. Current measurements do not show any clear evidences of the Fe/Mg evolution \citep{deR11}, however, the observed scatter at high redshift is large. 

It will be possible to observe high redshift clusters with {\it XRISM} with sufficiently long exposures ($\ge$ 100 ks). Feasibility studies showed that the Fe abundance will be measured with a $5\sigma$ significance with a 100 ks exposure out to $z=1$ clusters. Si abundance can be probed up to $z=0.6$, while O only feasible for a few systems up to $z=0.3-0.4$ \citep{Kit14}. 

Another way to probe high-$z$ chemical composition is to study absorption lines imprinted by the circumgalactic and intergalactic media in the spectra of bright background beacons such as the X-ray afterglows of gamma-ray bursts (GRBs) and distant blazars. {\it XRISM} will resolve these absorption features, allowing a study of the intervening warm-hot intergalactic medium (WHIM). {\it XRISM} will also detect  emission lines in GRB spectra, which are important for probing the high-$z$ environment of GRB progenitors \citep{Tas14}.

\subsection{Dark Matter Emission Lines}

Using stacked spectra from dozens of galaxy clusters at different redshifts, both {\it XMM-newton} and {\it Chandra} detected a weak unidentified emission line at around 3.55 $\pm 0.03$ keV (cluster frame)\citep{bulbul14}. A similar feature was found in {\it XMM-Newton}\/ spectra of M31 and the Perseus cluster\citep{boyarsky14}, and several later works reported detections from the Milky Way. While the nature of this feature is not known, an intriguing possibility is that it could be the decay line from sterile neutrinos --- a dark matter candidate --- with a mass of 7.1 keV. All these results relied on the low spectral resolution of CCD spectrometers. The \citep{bulbul14} result found a much stronger signal from the Perseus cluster compared to that from the stacked cluster sample. In 2016, {\it Hitomi}\/ observations of the Perseus cluster did not detect the line at the high flux reported by \citep{bulbul14}, though the statistics were limited \citep{hitomi17}. While faint line detections in any single source using a relatively poor CCD resolution are susceptible to systematic and calibration uncertainties (and the Perseus cluster detection may have been an example of that), the stacked-cluster signal\citep{bulbul14} is more robust with respect to the various instrumental effects. It has not been verified with high spectral resolution thus far. Combining all cluster observations with {\it XRISM}\/ over several years will offer the chance to detect this possible signal from dark matter decay.

\section{DIFFUSE GAS IN LOCAL ENVIRONMENTS}
{\it (Primary editors: L\'{i}a Corrales, Elisa Costantini})

\subsection{Missing Baryons and the WHIM}

Current observations of hot gas in galaxy clusters, the Local Group, and the warm intergalactic medium (IGM) indicate that ~20-40\% of the baryons seen in the high-z Universe are ?missing? from our observations at z=0 \citep{bregman07,shull12}. The missing baryons are suspected to occupy the filaments of gas between galaxies at temperatures $>$ 10$^{6}$ K: also known as the WHIM. The dominant ions with emission and absorption features in the X-ray band are O VII, O VIII, and Ne IX. Measurements of the O VII resonant absorption line at 21.6 \AA\ arising from the WHIM have been reported, but all to $<$ 5$\sigma$ threshold. Additionally, there is confusion as to whether the absorption seen so far corresponds to the IGM or to material within the Local Group \citep{bregman07}. Recent work gave evidence for O VII absorption in the bright X-ray blazar 1ES 1553+113, from z=0.43 and 0.35, offering a tentative detection for the hot WHIM that accounts for 9-40\% of the baryons\citep{nicastro18}. 

%%%%%%%%%%%%%%%%%%%%%%%%%%%%%%%%%%%%%%%%%%%%%%%%
\begin{figure}[htb]
    \centering
%    \vspace{16mm}
        \includegraphics[width=8cm]{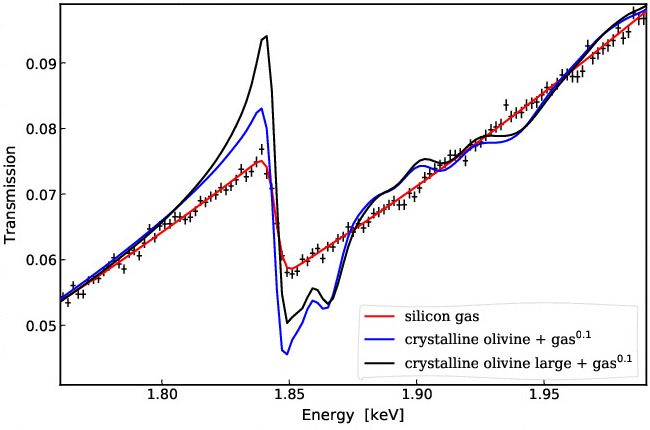}
        \includegraphics[width=8cm]{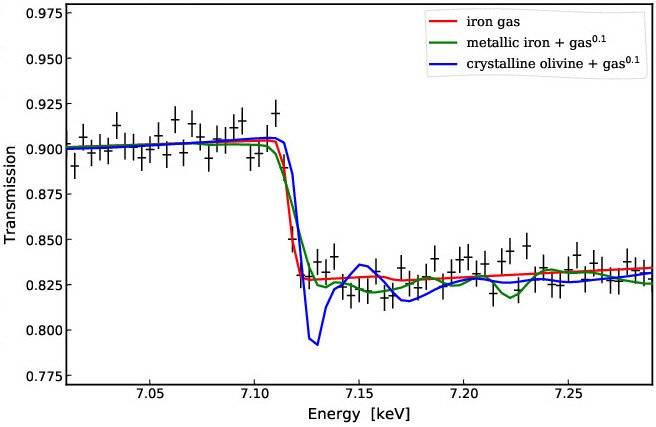}
    \caption[]{\footnotesize{{\it Left:} A 50 ks Resolve simulation of the Si K edge for a source with N$_{\rm H} = 2 \times 10^{22}$ cm$^{-2}$. {\it Right:} a 200 ks simulation of the Fe K edge for a source with N$_{\rm H} = 8.5 \times 10^{22}$ cm$^{-2}$.}}
%    \vspace{8mm}
\label{sikedge}
\end{figure}
%%%%%%%%%%%%%%%%%%%%%%%%%%%%%%%%%%%%%%%%%%%%%%%%

{\it XRISM} will study potential emission of the WHIM from the outskirts of more extended cluster regions (e.g., at distances more than two times the virial radius) in objects such as the Coma cluster \citep{finoguenov03}, Abell 85 \citep{durret03}, and Abell 2218 \citep{kaastra03,takei07}. {\it XRISM} will also measure the baryon fraction in gas-poor groups, big ellipticals or massive spirals, making it better able to characterize the thermal state and metal abundance of gas emission from lower mass structures in the Universe.

\subsection{Absorption in the ISM}

While only about 1\% of the ISM is condensed into solid dust particles, dust imprints unique extinction features in the spectra of background X-ray sources. The extinction features of Si, Mg, O, and Fe have been studied in detail with current X-ray missions \cite{zeegers17,zeegers19,costantini12,pinto13,valencic15,lee09}. These elements are the main constituents of silicates, which, together with carbon, dominate the chemical composition of interstellar dust. Through high-resolution X-ray spectroscopy, it has become clear that the X-ray band offers a privileged point of view on the properties of interstellar dust. 

The features caused by dust are due to the interaction between the incoming X-ray and the electrons inside the grain, which form an interference pattern, depending on the energy of electrons and the complexity of the compound. %Calculations of these patterns as well as laboratory measurements of dust allow us to interpret the astronomical spectra.
%We now know that silicates, mostly in the form of olivine (e.g. MgFeSiO$_{4}$), play a substantial role in the interstellar dust of moderately dense (N$_{\rm H} > 1 \times 10^{22}$ cm$^{-2}$) environments. Silicon in gas form contributes only about 10\% of the total absorption. 
These features are not only sensitive to the chemical composition of the intervening dust, but potentially to the grain size \citep{corrales16} and its crystallinity \citep{zeegers17,zeegers19}. 

Due to UV radiation and cosmic rays, dust is believed to undergo a process of amorphization. According to infrared studies, no more than 2\% of dust should survive in crystalline form. However, X-ray features are sensitive to short-range order in the grain, while the infrared features map long range order \citep{speck11}. This allows X-rays to be more sensitive to the crystalline fraction in a complex structure, where amorphous and crystalline lattice coexist. Recent X-ray studies suggest that the crystalline fraction could be up to 12\% \citep{zeegers19}.  The study of a single feature in the X-ray spectrum can reveal details about dust in different Galactic environments that are elusive at longer wavelengths.
 
{\it XRISM} will open a new parameter space with the investigation of the dense environments of the Galaxy (N$_{\rm H} = 1-10 \times 10^{22}$ cm$^{-2}$) near the Galactic Center or molecular clouds. Figure~\ref{sikedge} shows a 50 ks simulation of the spectrum in the Si K edge region (E $\sim 1.84$ keV), highlighting the impact of different grain composition and distributions. {\it XRISM} offers also the advantage of a negligible contamination by silicon in the detector itself. With {\it XRISM} we will be able to compare the spectral features of interstellar dust -- especially for heavier elements such as Ca, S, and Fe --  to material absorption measured in the lab. In addition, dust scattering produces a feature on the low-energy side of the absorption cross-section that results in an apparent spectral peak. The height of this peak directly measures the abundance of large grains ($> 0.3~\mu$m) and will allow {\it XRISM} to evaluate departures from the standard dust size distribution \citep{mathis77,draine09}.
% Old text:
%With XRISM we will be able to distinguish with clarity between amorphous and crystal silicates (orange and purple line) and between a standard dust size distribution \citep{mathis77} and a distribution skewed towards larger grains \citep{draine09}. The two distributions tested would be representative of a diffuse dust (radius = 0.05-0.25 $\mu$m) and of a more dense environment, respectively.  

%%%%%%%%%%%%%%%%%%%%%%%%%%%%%%%%%%%%%%%%%%%%%%%%
\begin{wrapfigure}{L}{0.55\textwidth}
    \centering
%    \vspace*{-5mm}
    \includegraphics[width=0.55\textwidth]{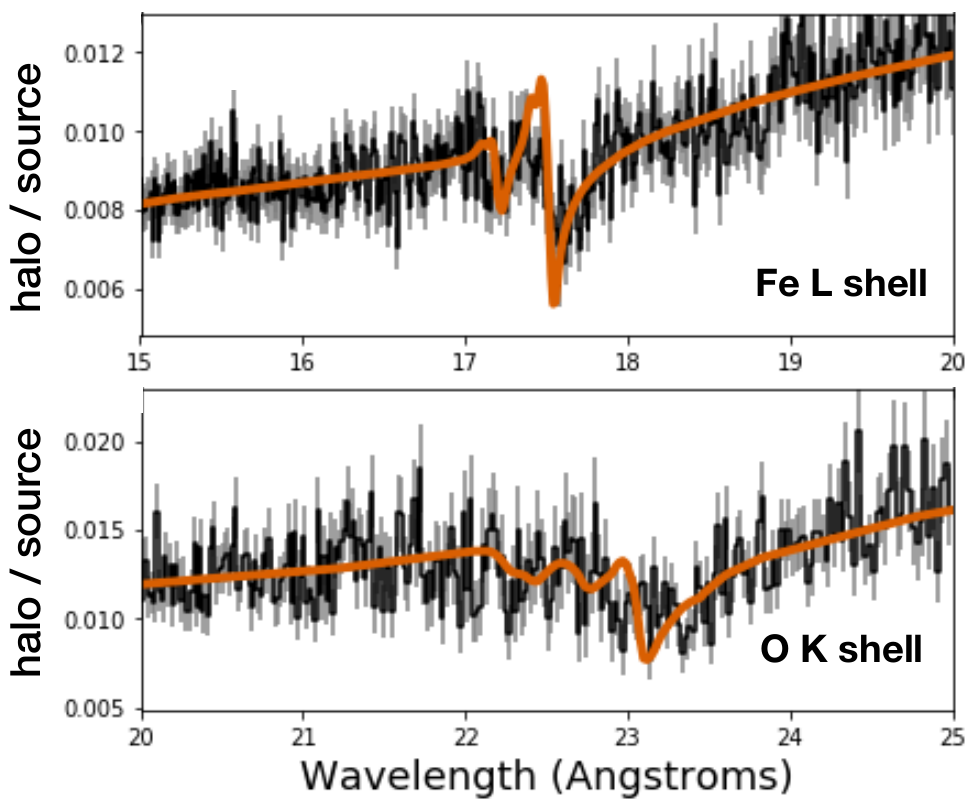}
    \caption[]{\footnotesize{Ratio of the scattering halo component to the point source  component.   The  X-ray scattering fine structure  models  for  Fe  L  and  O  K  (before applying the telescope response) are shown in orange.}}
\label{fig:dscatspectra}
\vspace*{-2mm}
\end{wrapfigure}
%%%%%%%%%%%%%%%%%%%%%%%%%%%%%%%%%%%%%%%%%%%%%%%%

In Figure~\ref{sikedge} we also show the simulation of the Fe K edge at 7.1 keV \citep{rogantini18}. The Fe K edge will be detected only for environments with N$_{\rm H}$ $>$ 8$\times$10$^{22}$ cm$^{-2}$. Persistently bright sources with a high column density are not frequent in our Galaxy, so the limiting factor in an accurate study of the feature will be the signal to noise ratio \citep{lee05}. {\it XRISM} will be able to distinguish between gas and dust absorption, offering a measure of depletion and abundance of iron in regions that are too absorbed to be observed at other wavelengths. 

Large column densities will increase the opacities of less abundant elements, like S and Ca\citep{costantini19}. In every ISM environment, Ca is completely depleted. It is believed that Ca, together with Al, is at the core of silicate grains. The silicate themselves would create a shielding mantle against radiation in the ISM, keeping the Ca and Al in dust form in every environment. {\it XRISM} will be able to identify absorption by silicates containing Ca and Al even if absorption by this element produces a transmission of only 1-2\%.

\subsection{Dust Scattering and Haloes}

Small angle forward scattering of X-ray light by interstellar dust grains leads to the phenomenon of dust scattering halos, an extended X-ray image spanning 10-20 arcminutes from a bright central source \citep{overbeck65,hayakawa70,mauche86}. When a compact object exhibits a bright X-ray outburst, this phenomenon manifests as bright rings of X-ray light that grow in angular size with time, referred to as dust ring echoes. The time and spectral information available with dust ring echoes allows for a deconvolution of the intervening dust location relative to the point source, the light curve of the outburst, and the dust grain size distribution \citep{heinz15,heinz16}. 

Like the high-resolution absorption features discussed previously, the scattering cross-section will also exhibit high-resolution signatures of dust composition. The spectrum of a dust scattering halo is also sensitive to the composition and elements bound up in dust grains of radius $\sim 0.3 - 3$ $\mu$m, which cannot be probed at any other wavelength. This may be related to the ``missing oxygen problem,'' which arises from the fact that along the most highly depleted sight-lines, more oxygen is observed to be missing from the gas phase than can be explained by its incorporation into silicates \citep{jenkins09}. This could be explained if oxygen is bound up in large micron-scale ice grains, which also prevents them from being detected via infrared spectroscopy. Due to shielding, large dust grains are also hidden from X-ray absorption \citep{wilms00}. Thus, high-resolution spectra of X-ray scattering halos in combination with high-resolution absorption features from the spectrum of the central point source are necessary for a full picture of interstellar dust composition (Figure~\ref{fig:dscatspectra}).

Resolve offers the opportunity to study the properties of dust in individual clouds in the ISM.  This arises from the fact that dust scattering halos and ring echo images exhibit asymmetries caused by either enhanced scattering from distant clouds \citep{mccollough13} or enhanced absorption from foreground clouds \citep{heinz15}. The Xtend instrument can be used to track the evolution of dust ring echoes, providing essential timing information that will identify the location of the interstellar clouds. {\it XRISM} will be an essential observatory for tracking processes of dust destruction and growth across various phases of the ISM, and will provide complementary datasets to the infrared spectra obtained by {\it JWST}. 

%%%%%%%%%%%%%%%%%%%%%%%%%%%%%%%%%%%%%%%%%%%%%%%%
\begin{wrapfigure}{R}{0.55\textwidth}
    \centering
    \vspace*{-3mm}
    \includegraphics[width=0.55\textwidth]{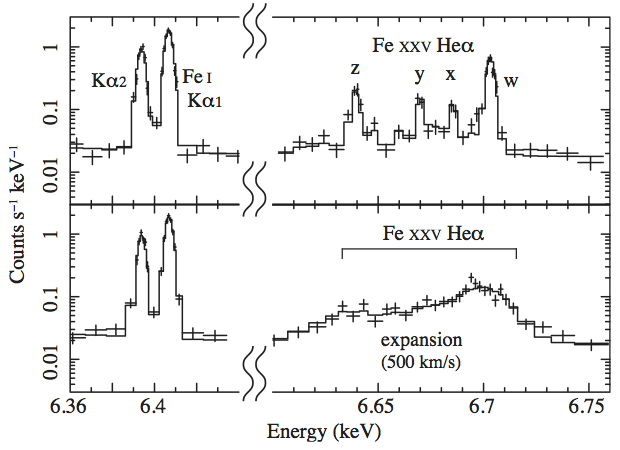}
    \caption[]{\footnotesize{Simulated spectra of the GCXE (the Radio Arc region). Upper panel is for the case of narrow Fe XXV K line (point source origin) line. Lower panel is same as the upper panel, but the Fe XXV K line is expanding with the velocity of 500 km/s (diffuse plasma origin\cite{koyama14}).}}
\label{galacticcenter}
%\vspace*{-5mm}
\end{wrapfigure}
%%%%%%%%%%%%%%%%%%%%%%%%%%%%%%%%%%%%%%%%%%%%%%%%

\subsection{The Galactic Center}

One of the most interesting phenomena in the center of the Milky Way is the Galactic Center X-ray Emission (GCXE). The GCXE consists of diffuse plasma with a spectrum characterized by strong atomic lines (Si to Ni, especially Fe). Only 10-40\% of the GCXE is resolved in point sources even by long-exposure {\it Chandra} observations \citep{muno03,revnivtsev07,zhu18}. Recently, {\it Suzaku} \citep{nobukawa16} revealed that the spectrum of the GCXE cannot be reconstructed by the assembly of well-known point sources (magnetic or non-magnetic cataclysmic variable and active binaries). In addition, {\it Chandra} \citep{zhu18} shows the equivalent widths of Fe K lines of the integrated spectrum of point sources in the GC region, and they are different from those of the GCXE shown by {\it Suzaku} \citep{yamauchi16,nobukawa16}. These facts indicate that the majority of the GCXE might come from truly diffuse hot plasma. 

With {\it XRISM}, we can measure the Doppler shift of and line broadening of the Fe XXV K line in the GCXE with accuracies of 45 km s$^{-1}$ and 300 km s$^{-1}$, respectively. If the majority of the GCXE is diffuse plasma of kT $\sim$ 7 keV (T $\sim 8 \times 10^{7}$ K), then it is not gravitationally bound to the GC region, and thus the rotation curve of the GCXE can be different from that in the radio band. In the case of a point source origin, narrow line broadening is expected, in contrast to expanding diffuse plasma\citep{koyama14} (see Figure~\ref{galacticcenter}).  
   
The non-thermal component characterized by the Fe I K$\alpha$ line is thought to be an X-ray echo from molecular clouds (MCs) irradiated by past X-rays from Sgr A* \citep{inui09}. The detection of the Compton shoulder of the Fe I K$\alpha$ line by {\it XRISM}  will constrain the positions and physical properties of the irradiated MCs \citep{odaka11}. On the other hand, {\it Suzaku} \citep{nobukawa15,nobukawa16} found that some fraction of Fe I K$\alpha$ in the Galactic disk region is not from the integration of faint point sources, but from low-energy (keV-MeV) cosmic rays (LECRs). It indicates that an LECR component would also exist in the GCXE. The high-resolution spectra of the GCXE might be able to resolve the unknown LECR components from the X-ray echo components and measure the amount in the GC region.

\subsection{Charge Exchange in the Solar System and Beyond}

With the first observations of comet Hyakutake with {\it ROSAT} in March 1996, astronomers were surprised to discover that the comet was 100 times brighter in X-rays than predicted \citep{lisse96}. It is now known \citep{cravens97} that the X-rays were produced primarily by charge exchange (CX) collisions between hot plasma ions from the solar wind and cold neutral atoms in the cometary atmosphere. This process is also observed when the solar wind interacts with neutral gas in the heliosphere \citep{robertson01} and planetary atmospheres \citep{cravens03,hui09}. CX can occur in any region in which a hot plasma interacts with a cold neutral gas, making it a useful diagnostic tool for studying astrophysical plasmas and the ISM. Heliospheric and exospheric foreground CX emission also poses an ?astrophysical nuisance?, as it contaminates all X-ray observations \citep{kuntz19}. Foreground CX emission particularly affects diagnostics of the X-ray emission from the local bubble \citep{galeazzi14}. CX is also expected to contribute to the X-ray spectra from astrophysical sources outside of our solar system, such as in galaxy clusters \citep{lallement04}, supernova remnants \citep{katsuda11}, and star forming galaxies \citep{zhang14,konami11}. CX emission will be discussed further in Section~\ref{stars}.

%During a CX collision, an ion captures one or more electrons from a neutral atom or molecule,
%producing an ion in a highly excited state which will emit EUV and X-ray photons as the captured electron(s) cascade(s) to the ground state. The quantum state-selective cross-sections and thus the resulting X-ray emission lines for a CX collision vary significantly depending on the ion and neutral species as well as the velocity of the collision. The spectrum resulting from CX is very different than that from thermal emission, as highlighted in Figure~\ref{fig:cx}. Spectroscopy is thus highly diagnostic of both the importance of CX as well as the physical conditions of the CX interaction.
 
{\it XRISM} will readily discriminate CX emission mechanisms from other processes, and provide better constraints for CX within our own solar system. Observations of cometary spectra will help constrain the neutral density in a cometary coma and the solar wind ion abundance, and observations of the Mars and Jovian atmospheres will place constraints on the importance of CX in these atmospheres, helping us understand the relative significance of other processes producing X-rays.  

%***************************
\begin{comment}
\begin{figure}[htb]
%vspace{-0.2in}
\begin{center}
\includegraphics[width=1.0\textwidth]{vel_uncert.png}
\end{center}
\vspace{-3mm}
\caption[]{\footnotesize {Combined statistical and systematic uncertainties on turbulent (left) and bulk velocities (right) in {\it XRISM} simulations as a function of the amplitude of turbulent motions. Systematic uncertainties are associated with calibration of the line spread function and gain \citep{Kit14}. For comparison, current CCD imaging spectrometers can ``see" only high-velocity motions with typical velocities $>1000$ km s$^{-1}$.}}
\label{fig:vel_uncert}
\vspace{-3mm}
\end{figure}
\end{comment}
%***************************

\section{PHYSICS AND DYNAMICS OF COSMIC PLASMAS}
{\it (Primary editors: Maxim Markevitch, Irina Zhuravleva)}\\

The ability to measure the widths and positions of emission lines of multiple elements at high precision will allow {\it XRISM} to perform unique diagnostics of the optically-thin plasma in a multitude of astrophysical systems. Below we discuss several avenues for such studies, which fall into two broad classes --- the dynamics based on line Doppler shifts and broadening, and the relative line fluxes. While Resolve is spending long exposures performing spectroscopic studies of the cluster bright central regions, Xtend, with its wide field of view and low instrument background, will obtain high-statistics maps of the plasma density, temperature, and abundances of main elements at much greater off-center distances, providing information on large-scale cluster dynamics that drive many of those spectroscopic phenomena. 

\subsection{Velocity Field of Intracluster Plasma} 

The hot gas in galaxy clusters is continuously stirred by mergers, feedback from AGN, motions of galaxies, magneto-thermal and plasma instabilities, etc. All these processes generate bulk and turbulent motions of the gas in the ICM. %Sharp surface brightness discontinuities associated with cold fronts \citep[e.g.,][]{Ett00,Vik01,ZuH11,Wer16,Wal18}, extended ram-pressure-stripped gaseous tails of galaxies infalling and moving through the ICM \citep[e.g.,][]{Roe15,Kra17,Su17}, and the absence of exponential steepening of the power spectrum of density fluctuations close to the particle mean free path scale \citep{Zhu19} point to significantly suppressed transport processes (in particular, gas viscosity and thermal conduction) in the intracluster plasma compared to the Spitzer value. This means that turbulence can easily develop in these environments. 
Turbulent motions cascade down towards dissipation scales, amplifying and tangling the ICM magnetic field \citep{Vaz14,Por15} and accelerating cosmic rays \citep{Fuj03,Bru07,Min15} through the second-order Fermi process. Using high-resolution {\it XRISM} spectra, we will map the gas velocity field in galaxies, clusters, and other extended X-ray sources with the accuracy of $< 100$ km s$^{-1}$, bringing a new dimension in the study of astrophysical plasmas. 

\begin{figure*}[htb]
%vspace{-0.2in}
\begin{center}
\includegraphics[width=1.0\textwidth]{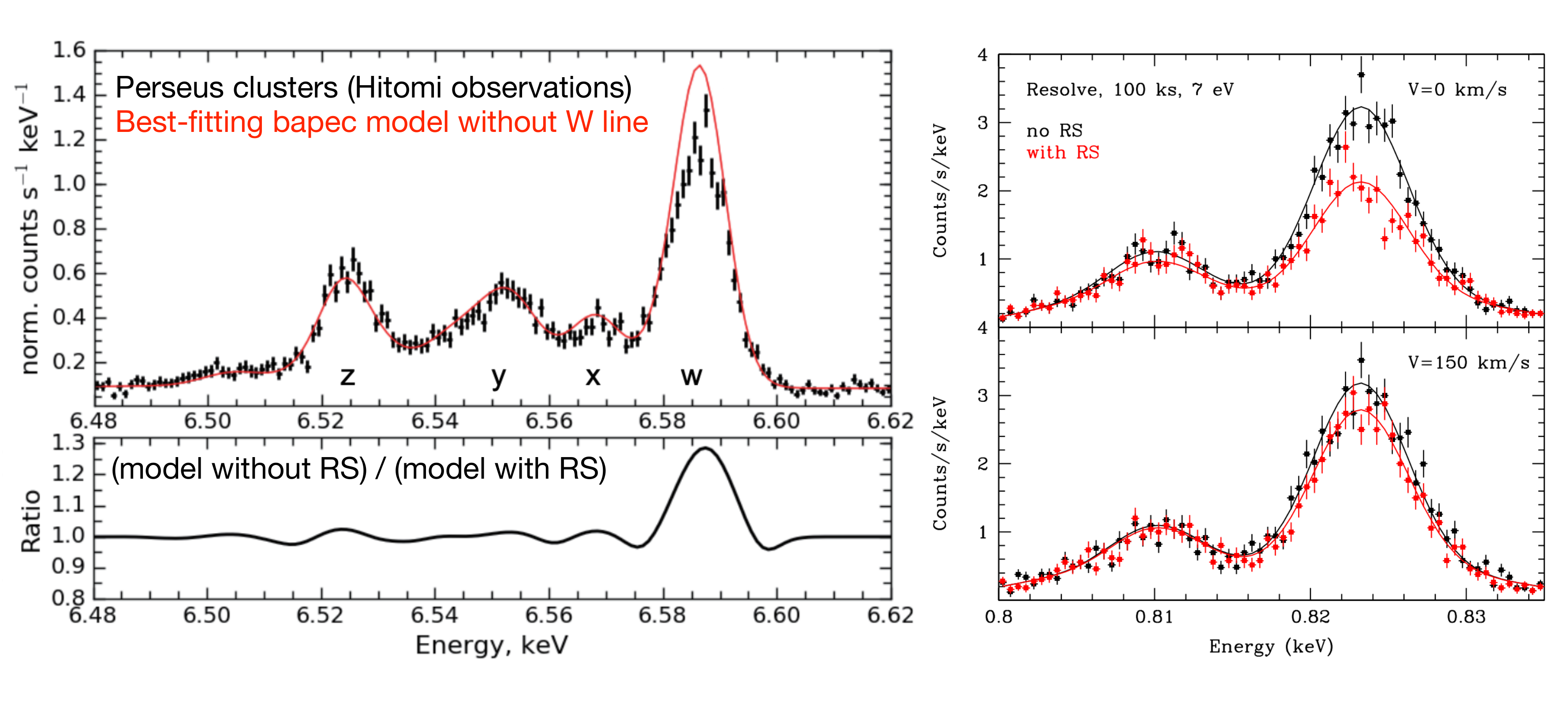}
\end{center}
\vspace{-0.2in}
\caption[]{\footnotesize {{\it Left:} {\it Hitomi} observations of resonant scattering in the core of the Perseus cluster. The strongest line of Fe~XXV (w) is suppressed by $\sim 20$\% in the center relative to the predictions for the optically-thin plasma\cite{HitRS}. {\it Right:} 100 ks {\it XRISM} simulated spectrum around the Fe XVII (0.826 keV) line from the center of NGC 4636, assuming no motions of the gas (top) and isotropic turbulence with V=150 km/s (bottom). Red and black lines: models with and without resonant scattering\cite{Ogo17}.}}
\label{fig:rs_hit_XRISM}
\vspace{-3mm}
\end{figure*}

{\bf Resonance Scattering in Strong Emission Lines} The same data used for the Doppler broadening measurements will provide complementary velocity constraints through the resonant scattering effect \citep{Gil87}. Resonant scattering reduces fluxes in some strong resonant lines in the cores of galaxies and clusters with respect to the fluxes in an optically-thin plasma. The degree of the suppression depends mainly on the amplitude of turbulent motions of the gas --- the stronger the motion, the smaller the effect. Therefore, the suppression of lines can be used as a powerful independent probe of small-scale motions \citep{Chu10,Gu18}. Such measurements have been successfully done in the innermost regions of galaxies and groups using high-resolution \xmm\ RGS data \citep{deP12,Wer10,Ogo17} as well as \Hitomi\ data \citep{HitRS} on Perseus (Figure~\ref{fig:rs_hit_XRISM}). Furthermore, by measuring the Doppler broadening and resonant scattering as a function of projected distance from the center, as well as the shapes of strong emission lines, one can place constraints on the anisotropy of gas motions \citep{Zhu11} in bright cluster cores. %This would require relatively long {\it XRISM} observations and complementary numerical simulations. The primary targets for the resonant scattering analysis include several bright galaxy clusters (Perseus, Virgo, Centaurus, mainly Fe K lines) and the brightest elliptical galaxies and groups (NGC4636, NGC1404, NGC4374, NGC5044).

%%%%%%%%%%%%%%%%%%%%%%%%%%%%%%%%%%%%%%%%%%%
%%%%%%%%%%%%%%%%%%%%%%%%%%%%%%%%%%%%%%%%%%%%

The strength of the resonant scattering depends on the line optical depth. For hot systems such as Perseus (T $\sim5$ keV), the effect is strongest for the resonant He-like Fe line at 6.7 keV ($\tau \sim 3$ if $v=0$). For cooler systems such as Virgo (T $\sim$ 2-3 keV), the thickest lines, $\tau\sim 1-1.5$, are Fe~XXIII ($E=1.129$ keV), Fe~XXIV (1.168 keV) and Fe~XXV (6.7 keV). In cool systems such as giant elliptical galaxies and groups (T $\sim 1$ keV), the resonant scattering effect could be even stronger since the optical depth of the strongest Fe XVII line (0.87 keV) could be up to $\sim 8-10$ (Figure~\ref{fig:rs_hit_XRISM}). 

{\bf Power spectrum of turbulence} For many astrophysical applications, it is important to understand spatial scales associated with the measured velocities of gas motions as well as statistical properties of turbulence (e.g., power spectrum or structure function). For instance, the turbulent heating rate, which is important for understanding energy transport in the ICM, is $\propto v^3/l$, where $v$ is velocity associated with turbulent motions on a scale $l$. The slope of the power spectrum within inertial range tells us about the intrinsic properties of turbulence (e.g., hydrodynamic vs. magnetohydrodynamic, isotropic vs. anisotropic) and transport processes in the intracluster plasma \citep{Zhu19}. 

\begin{figure*}[htb]
%vspace{-0.2in}
\begin{center}
\includegraphics[width=1.0\textwidth]{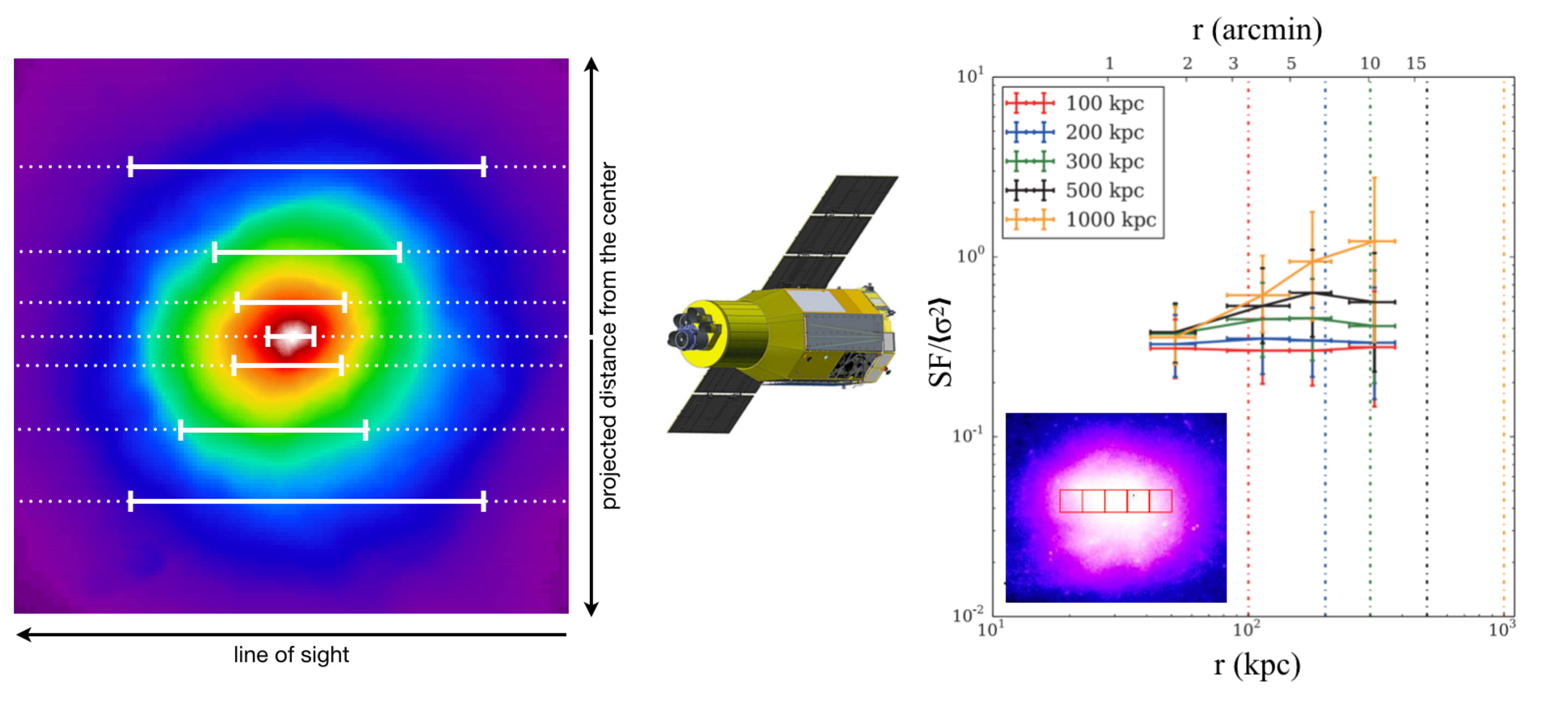}
\end{center}
\vspace{-0.2in}
\caption[]{\footnotesize {{\it Left:} In a cluster with a centrally peaked surface brightness, the projected radius $r$ determines the size of the region along the line of sight, from which the largest fraction of photons contributes to the line flux. This size corresponds to the dominant scale of motions we will probe at each $r$\citep{Zhu12}. {\it Right:} structure function of turbulence in the Coma cluster as {\it XRISM} would measure if spectra are measured in five pointing regions shown in the bottom left corner. The assumed Mach number of turbulence is 0.2. Different colors correspond to different assumptions about the injection scale of turbulence\cite{ZuH16}.}}
\label{fig:power_sp}
\vspace{-2mm}
\end{figure*}

Measuring the power spectrum of turbulence from the observed characteristics (line widths and shifts) is not a trivial matter. Some ideas tailored for galaxy clusters have been proposed \citep{Zhu12,ZuH16}. One of them utilizes a steep decline of the X-ray surface brightness with radius observed in cool-core clusters. In such clusters, the measured line broadening as a function of projected distance from the center would effectively probe velocities on different spatial scales (Figure~\ref{fig:power_sp}, left panel). For clusters with flat surface brightness profiles, such as Coma, mapping of the projected velocity field provides the most direct way of obtaining the velocity power spectrum \citep{Zhu12}. \cite{ZuH16} showed that even with a sparse coverage of the Coma cluster, {\it XRISM} will be able to determine the injection scale of the turbulence (Figure~\ref{fig:power_sp}). Although it is likely that these measurements will be done in a small sample of clusters, they will significantly advance the field and calibrate the density fluctuations - velocity relation that is currently used as an indirect probe for the turbulence power spectrum \citep{Zhu14a,Gas14}. %It is important to note that these techniques are approximations that do not take into account the effects of stratification and anisotropy of turbulence, multiple injection scales, and intermittency. Hydrodynamic simulations will be required for the physical interpretation of the data, possibly in combination with complementary data on resonant scattering. 

\subsection{Shocks in Clusters and Elsewhere}

Shock fronts are ubiquitous on all scales across the Universe, from the solar wind to supernova remnants (SNRs) to the intergalactic plasma. They convert kinetic energy into thermal energy of the medium in which they propagate and accelerate cosmic ray particles via the Fermi mechanism. The physics of shocks in magnetized astrophysical plasmas is uncertain and complex. {\it XRISM} will uncover interesting aspects of the shock physics in clusters and SNR. For clusters with large shock fronts, such as A754 and Coma, {\it XRISM}'s angular resolution is sufficient to isolate the post-shock region and derive its spectrum. {\it XRISM} will characterize the small-scale turbulence that could be driven by the shock and look for transient deviations from thermal and ionization equilibrium caused by the shock passage, probing the energy conversion mechanisms and the various equilibration timescales in the plasma. 

\subsection{Plasma Diagnostics}

While our usual approach of analyzing an X-ray spectrum is brute-force fitting of the whole energy interval with a physical model that predicts all the lines and continuum, in many cases the quantities of interest can be separated and determined independently using certain emission lines sensitive to those quantities. This can be particularly helpful when deviations from thermal and ionization equilibrium are possible or when the model includes nonthermal emission. {\it XRISM} will provide us with these new plasma diagnostics. 

When a plasma is in thermal equilibrium, all particle species (electrons, protons, and heavier ions) must have the same temperature, which means that ions of different atomic mass should have different thermal velocities. On the other hand, if the plasma has any bulk streams or turbulence, all ions participate in those motions at the same velocity. By measuring the line widths for different elements, we can disentangle the two broadening mechanisms and thus determine the ion temperature (Kitayama et al.\ 2014). Various interesting processes can result in temporary electron-ion nonequilibrium, such as a shock passage and  waves in the plasma. An attempt has been made to determine $T_i$ in Perseus \citep{hitomi18b}, which showed $T_i=T_e$ but with a large uncertainty. {\it XRISM} will make such measurements possible in many clusters and other astrophysical systems.

%%%%%%%%%%%%%%%%%%%%%%%%%%%%%%%%%%%%%%%%%%%%%%%%
\begin{wrapfigure}{R}{0.55\textwidth}
    \begin{center}
    \vspace*{-5mm}
    \includegraphics[width=0.42\textwidth,angle=90]{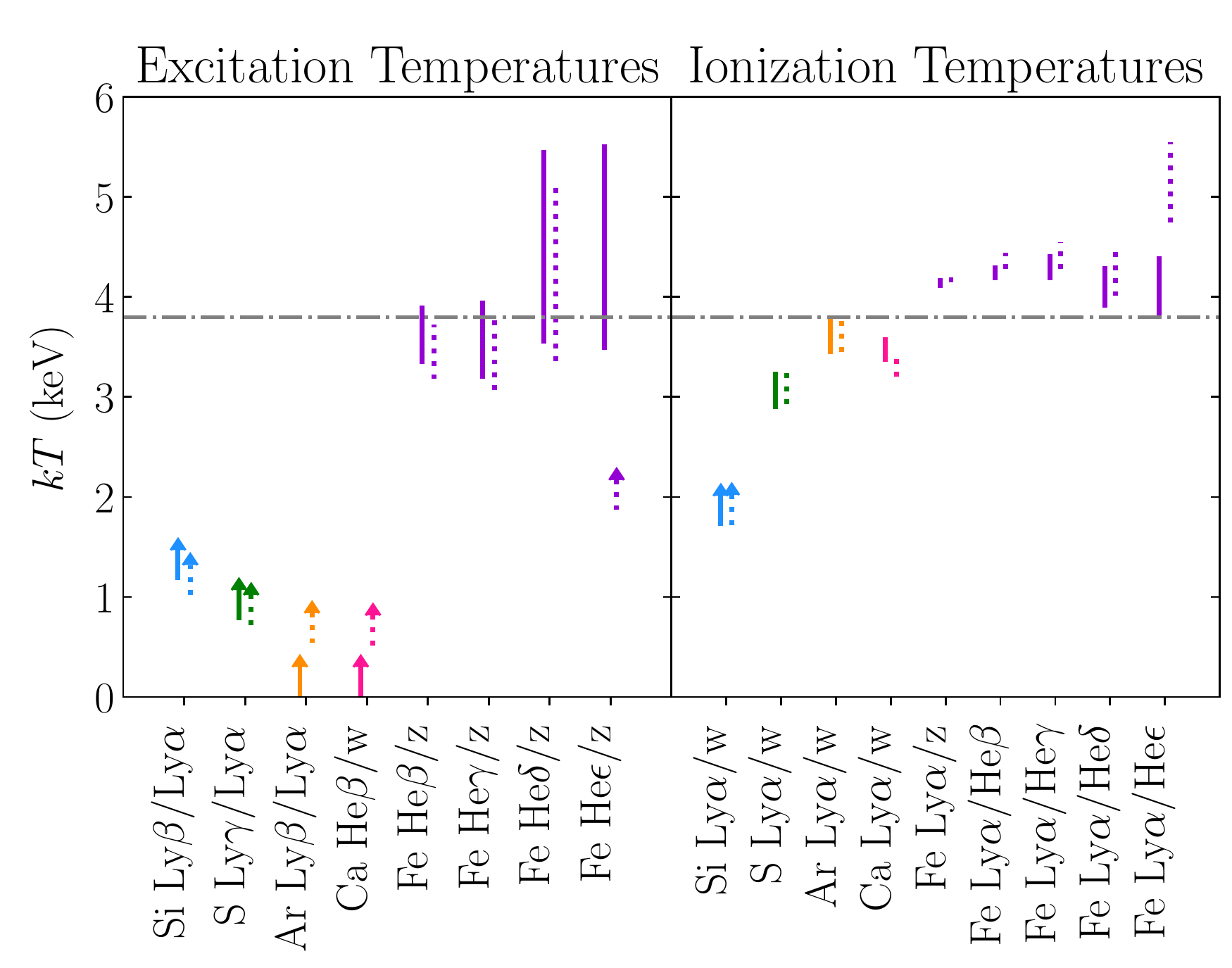}
    \end{center}
    \caption{\footnotesize {Line ratios from the \hitomi\ spectrum of Perseus \citep{hitomi_temps18}; labels {\em w}\/ and {\em z}\/ refer to resonant and forbidden transitions of the He$\alpha$ complex. {\em Left}: $T_e$ derived from pairs of lines for the same ion. {\em Right}: the ``ionization temperature'' for each element, derived from the lines for different ions of the same element; their significant differences may indicate a deviation from ionization equilibrium or a mixture of temperatures. Solid and dashed lines are for different atomic codes.}}
\label{fig:lineratios}
%\vspace*{-5mm}
\end{wrapfigure}
%%%%%%%%%%%%%%%%%%%%%%%%%%%%%%%%%%%%%%%%%%%%%%%%

In a plasma that has been rapidly heated (e.g., by a shock), different ions take different amounts of time to arrive at the ionization state determined by the electron temperature, and for a short period, the plasma can be out of ionization equilibrium \citep{Inoue16}. This effect, common in SNRs, manifests itself as an anomalous ratio of the abundances of different ions of the same element with respect to that expected for the measured $T_e$ under the equilibrium assumption --- it would correspond to a different ``ionization temperature." In clusters, the ionization equilibrium timescales are generally very short, so the detection of any such deviation can indicate a very recent heating event. For a system that is a mixture of plasmas at different temperatures (in 3D or in projection), lines from different elements ``highlight'' different temperature phases. 

The relative population of different excitation states for the same ion reflects the temperature of free electrons in the plasma. Thus, the ratio of lines emitted by the same ion can give an independent measure of $T_e$, complementary to that determined from the shape of the X-ray continuum. This is especially useful when there is a mixture of plasmas at different temperatures along the line of sight --- as is often the case for extended sources. The \Hitomi\ Perseus observation has been used to determine the ionization temperature and $T_e$ from the line ratios (the ``excitation temperature'') using several line pairs (Figure~\ref{fig:lineratios}, \citep{hitomi_temps18}). The excitation and ionization temperatures are consistent with a mixture of temperatures in a relatively narrow range. {\it XRISM} will make these temperature diagnostics available for many clusters.

The thermal plasma in galaxy clusters is permeated with magnetic fields and ultrarelativistic ($\gamma\sim 10^4$) electrons, which reveal themselves via radio synchrotron radiation \citep{vanweeren19}. While these components contribute little to the ICM pressure, they can alter its microscopic properties quite significantly. The radio synchrotron measurements cannot disentangle the magnetic pressure from the density of the relativistic electrons, but Inverse Compton (IC) emission from the same electrons --- upscattering of the Cosmic Microwave Background photons into the X-ray band --- can \citep{harris74}. The X-ray IC spectrum should be a power law that can be deduced from the radio synchrotron emission. Searches of such power-law emission from clusters have been numerous but unsuccessful, because its surface brightness is far lower than the thermal emission and because the ICM is multi-temperature, so any ``hard tail'' in the residuals of a fit of the continuum by a thermal model can be interpreted as another thermal component \citep{wik14}. The latter complication can be alleviated by deriving the temperature structure of the plasma from the spectral lines (see above) rather than the continuum fits, because a hot thermal component that is needed to explain a hard continuum excess must also produce emission lines characteristic of that temperature.

\section{FORMATION AND EVOLUTION OF COMPACT OBJECTS}
{\it (Primary editors: Jon Miller, Chris Done, Mar\'{i}a D\'{i}az-Trigo})

\subsection{Winds from Compact Galactic Sources}

Winds are ubiquitous from accretion disks, and are detected in black hole and accreting neutron star
binaries with low magnetic fields \citep{ueda04,miller06a,miller06b,diaztrigo07,neilsen09,king12,king15,ponti12,miller16a,zoghbi16}, and some pulsars \citep{reynolds10,miller11,degenaar14}.  They are also detected in
accreting white dwarf systems, predominantly in UV and EUV spectra \citep{mauche87,mauche00,long96,long03,wheatley05}. Disk winds and outflows are more widely studied in AGN, where such flows can potentially control the star
formation-powered growth of the host galaxy.  However, understanding
winds and jets from accretion onto stellar-mass compact objects can
provide an independent angle on the nature of winds and jets from
massive black holes \citep{king13}.

Indeed, stellar-mass compact objects offer some unique
opportunities.  Whereas an AGN lifetime may typically be $\sim10^{7}$~
years \citep{martini04}, mass accretion rates can change by orders
of magnitude over just weeks and months in stellar-mass compact
objects \citep[]{remillard06,reynolds13}.
Moreover, although the kinetic power of winds from stellar-mass
compact objects can be modest, their role and impact may be anything
but.  They may be a manifestation of the processes that drive
mass and angular momentum transfer within the accretion disk, enabling
observational tests of analytical and numerical accretion disk theory
\citep{miller15a,miller16b}.  In some cases, winds (and nova
episodes in white dwarf binaries) may expel more gas than the system is able to
accrete, altering the evolution of the binary system and compact
object \citep{king12,king15}.  High-resolution X-ray
spectroscopy is the key to learning how these winds are launched
(which also determines how much energy, momentum and mass they carry),
and how they connect to the accretion flow and its relativistic radio
jet \citep{fukumura17,higginbottom17,done18,waters18}.

{\bf Disk-dominated states} Winds are seen when the accretion flow is dominated by an accretion
disk: the high/soft state in black hole X-ray binaries, the ``banana
branch" in typical neutron star X-ray binaries, and outbursting or
dwarf nova white dwarf systems \citep[]{miller06a,neilsen09,king12,ponti12}.  Winds largely
disappear when the source makes a transition to spectrally harder
states and/or lower Eddington fractions.  Typically, a radio jet appears
as the source makes this transition \citep{fender04}.  This general anti-correlation of the jet and the wind has led
to some speculation that these flows are fundamentally part of the same
processes, with the state change leading to a re-organization of the
magnetic field from powering the jet to powering the wind \citep{livio03,begelman15}.
Alternatively, the response of the wind could instead be linked to the
change in the illuminating spectrum at the state transition, with
over-ionization leading to a lack of atomic transitions.  Indeed,
these factors are not mutually exclusive and may act concurrently.

Non-magnetic wind models are either radiation-driven (for $L\geq
L_{Edd}$), thermally driven, or UV line driven.  The disks around
white dwarfs are bright in UV, and the corresponding winds have a
modest ionization; UV line driving is likely the dominant process by
which winds are expelled from these disks.  In contrast, black hole
X-ray binary and neutron star X-ray binary disks are too hot for UV
line driving to be important, and the relative importance of thermal
and magnetic processes in driving winds in these systems remains
unclear.  The ability of high-resolution spectroscopy to accurately
measure velocities, to separate distinct wind zones, and to measure
launching radii will be crucial to resolving their nature.

\begin{figure}
\includegraphics[width=8.5cm]{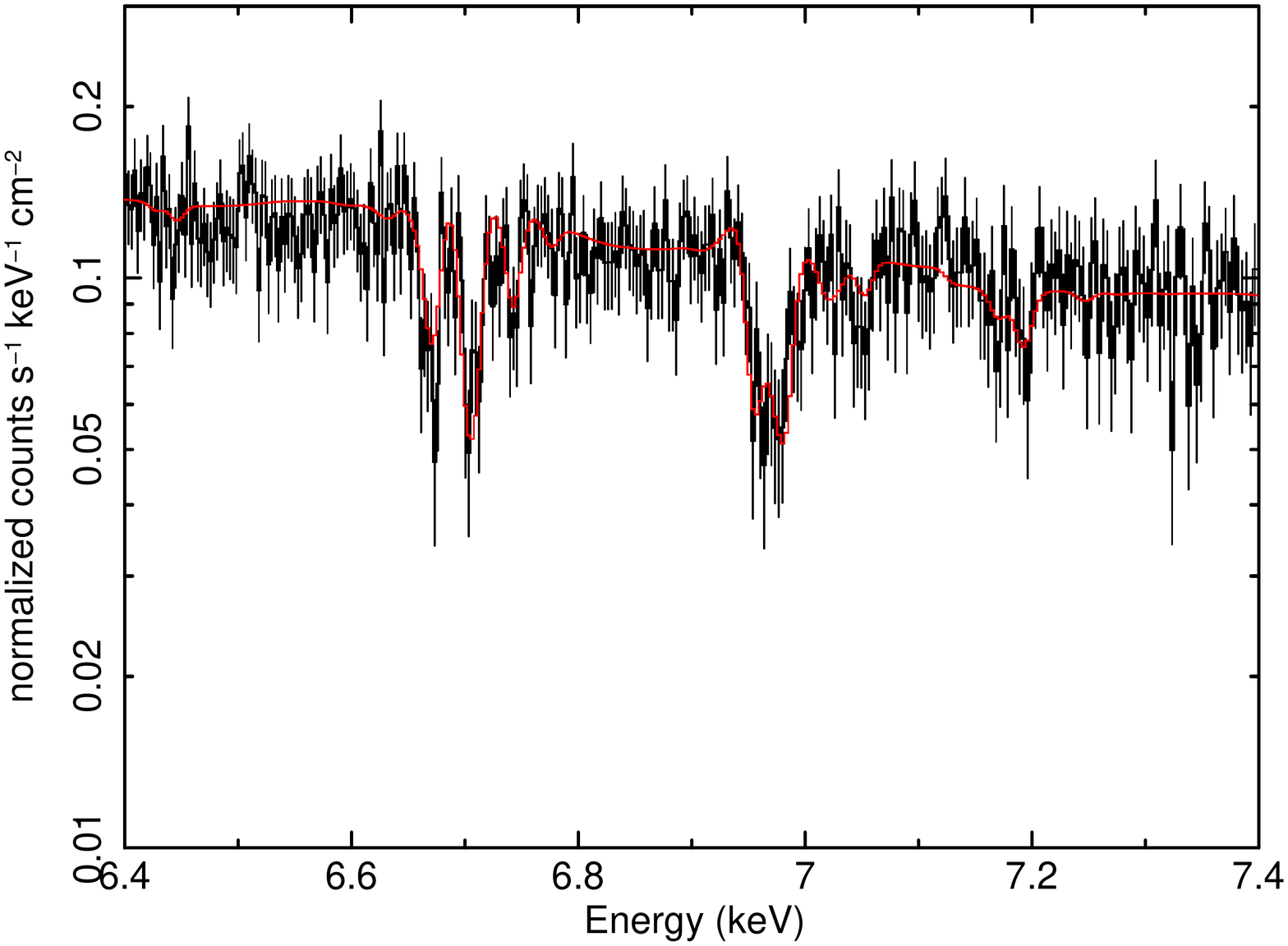}
\includegraphics[width=8.5cm]{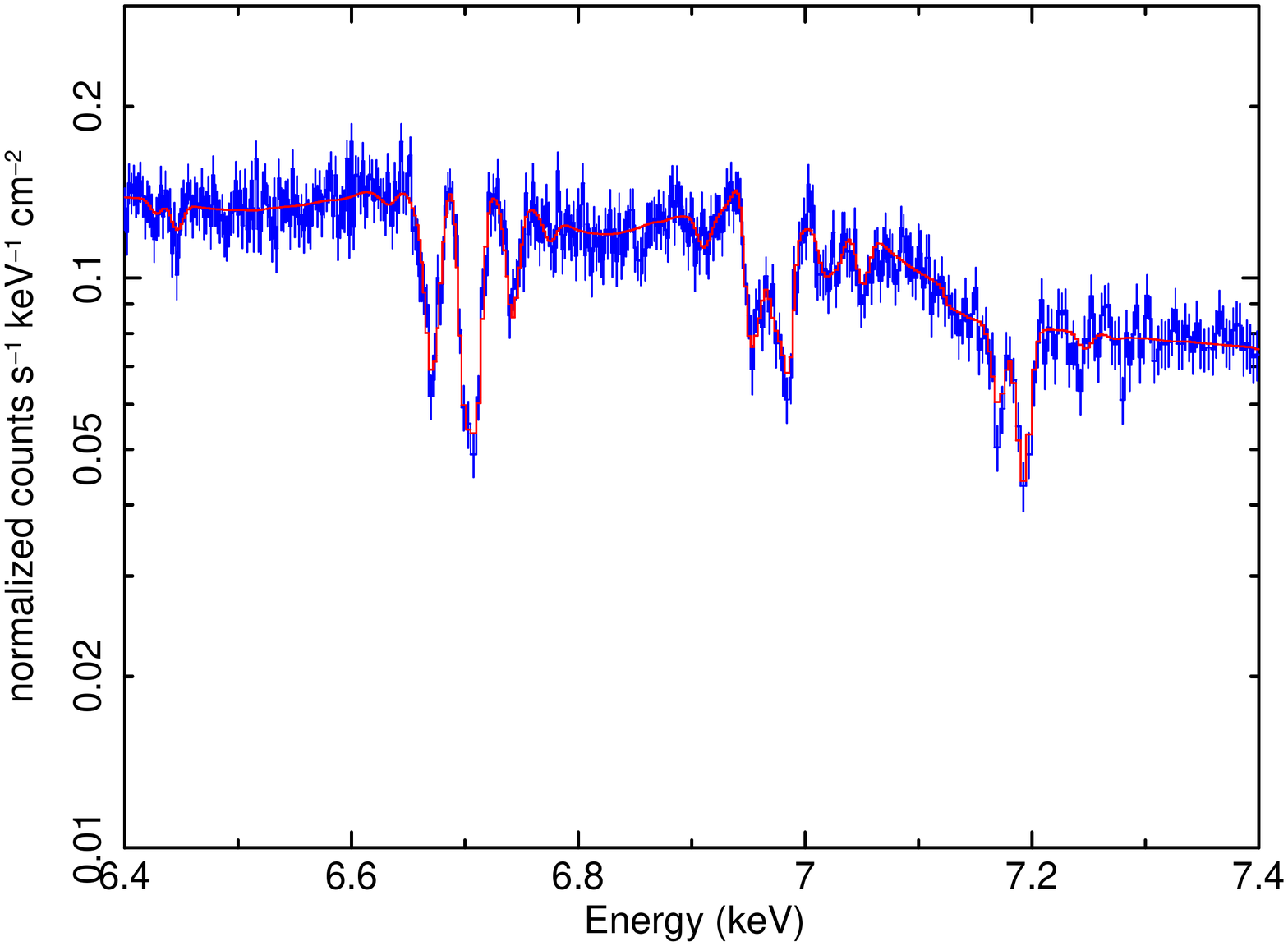}
\caption{\footnotesize {\it Left}: The {\it Chandra}/HETG third-order
  spectrum of GRS 1915$+$105\cite{miller16b}. The Fe XXVI
  line is revealed as a spin-orbit doublet splity by 22~eV, and four
  separate absorption zones are required.  The fastest zones have
  blueshifts of 0.01 and 0.03c, and are evident just above 7.0 and
  7.2 keV. {\it Right}: A simulated 50 ks Resolve spectrum of GRS 1915$+$105 in
  the same state, using the best-fit model of the HETG. In the simulated data, re-emission that can be used to measure
  launching radii is easily detected just to the red of the absorption
  lines, as are the fastest, most powerful wind components.}
  \label{ch4fig1}
  \vspace{-3mm}
\end{figure}

The excellent spectral resolution of Resolve will
make it possible to separate complexes of lines, and to examine the
velocity profile of specific strong lines. Density-sensitive lines from intermediate charge states \citep{mauche00,miller06b,miller08,mao17} will be
accessible to Resolve, by virtue of its broad pass band.  When the density of a
wind component is measured directly, the absorption radius can be
derived via the photoionization parameter.  Independently, velocity
broadening of the emission component in accretion disk P Cygni
profiles can be used to measure wind production radii \citep{miller15a,miller16b} (see Figure~\ref{ch4fig1}). 

The relative importance of emission and absorption features can determine the geometry of disk
winds, and trace the scale height of winds as a function of Eddington
fraction. Narrow emission lines can potentially reveal the scale height of the
outer disk, the extent to which it is flared, and how such structure
and ionization may relate to sharp flux dips \citep{diaztrigo06}.  Line broadening can reveal the characteristic radius,
while line flux is related to the fraction of the sky that is
subtended by the outer disk relative to the source.

{\bf Non-thermal X-ray states} There is a major spectral transition seen in all types of disk accreting systems, where the disk emission drops dramatically and the
non-thermal X-ray emission increases.  In black hole X-ray binaries
this is called the ``low/hard" state; in neutron stars, it is the
``island" or ``extreme island'' state; in dwarf novae, an equivalent
state is called quiescence (note, however, in this case the X-rays are
optically-thin thermal emission, not non-thermal). In all systems,
this transition correlates with stronger radio emission from the jet
\citep{fender04,remillard06}. There is still considerable debate concerning the origin and
nature of this transition, but one model is that the inner,
geometrically-thin, optically-thick accretion disk transitions into a
hot, geometrically-thick, optically-thin flow \citep{narayan94,esin97}.  Some theoretical treatments
suggest that an optically-thin accretion flow with a large scale
height is more efficient at producing a jet because poloidal fields
are easier to maintain in such a flow \citep{blandford77,mckinney12}.

Exactly why winds appear to be quenched or diminished in non-thermal
states is a complex problem.  Clearly, changes in the accretion disk
magnetic field topology, the geometry of the inner accretion flow, and
the spectrum emitted by the inner accretion flow may be closely
linked.  The exact sequence of such changes is not clear; it may
depend on, e.g., whether the mass accretion rate is increasing or
decreasing, or a number of other factors. Disk reflection and wind studies 
with Resolve will allow substantial progress. Winds might persist, but at lower column densities and
much higher ionizations; the unprecedented sensitivity and resolution
of Resolve can detect such winds.  Resolve will also make it possible
to separately trace the evolution of outflows, the inner disk, and the
outer disk, potentially revealing the importance of geometric changes
in diminishing outflows.

%\noindent $\bullet$ {\it What is the nature of the inner accretion
%flow in white dwarf systems?}  Separating reflection from the white
%dwarf surface and disk is difficult, but recent work has established
%its presence \citep{mukai15}.  The spectral resolution
%afforded by the Resolve calorimeter can help to separate reflection
%from the stellar surface and the disk.  Whereas emission from the
%stellar surface will have a narrow range in velocity broadening,
%emission from the disk will have a larger range (even if it is
%truncated), and any associated lines will be broadened.

\subsection{Surface Redshifts from Massive White Dwarfs}

As noted in the Astro-H white paper on white dwarfs \citep{mukai14}, a microcalorimeter may be able to directly
measure the redshift from the surface of a white dwarf.  The
anticipated shift is small (a few parts in ten thousand); however,
the centroid of a strong line can be measured to an accuracy that is
much sharper than a single spectral bin, particularly with the inclusion of the calibration pixel.

The motivations for this measurement are simple, but profound. It is
not clear how effectively white dwarfs can gain mass through
accretion; it may be that nova episodes -- with temporary nuclear
burning on the white dwarf surface -- serve to expel more gas than they accrete between such episodes. Finding a massive white dwarf
in a binary would clearly signal that accretion does add mass and can
thereby lead to SNe Ia, with consequences for elemental seeding within
galaxies and the use of SNe Ia as standard candles for cosmology.

In principle, the measurement is simple: the redshift of an Fe K$\alpha$ line
excited on the surface of the white dwarf determines $M/R$; since the
equation of state is known, this can be translated to $M$.  In
practice, there are several difficulties.  The Fe K$\alpha$ line is
actually a doublet, separated by 13 eV, and it may be the case
that a range of charge states are observed; Fe I-XVII lines all lie
within a range of just 30 eV. The motions of the bodies within
the binary system and the rotation of the white dwarf may add complications.
Finally, lines may not be excited only on the surface of the white
dwarf, though lines from the surface should be stronger than lines
from elsewhere in the system.

\subsection{High-mass X-ray Binaries}

The compact objects in most high-mass X-ray binaries likely accrete by
gravitationally focusing the wind driven by their massive companion
star.  Close to the compact object, the wind may be corralled into a
small accretion disk.  The exact geometry of the flow within the system
is complex; the accreting gas may be shocked, photoionized, and
collisionally ionized at different points.  The flow is not
axisymmetric, and may change considerably with source state.  In
pulsars such as Vela X-1, the pulse can act as a lighthouse that
sweeps across the accretion flow, helping to reveal its nature.  This
effect, combined with the orbit of the binary, especially in eclipsing
sources \citep{torrejon15a}, can potentially be utilized to study
the wind geometry in detail.

\begin{figure}[htb]
\includegraphics[width=8.5cm]{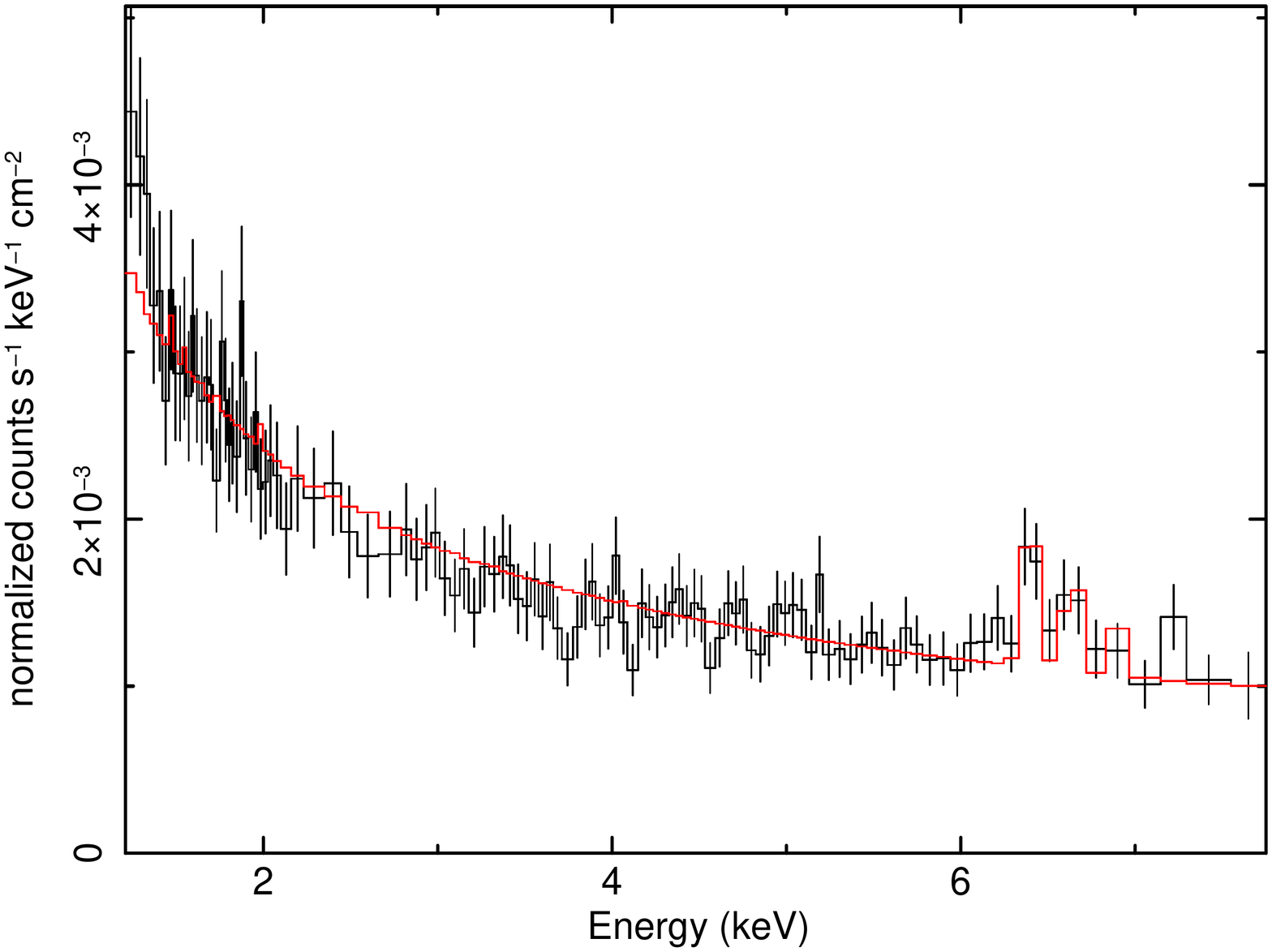}
\includegraphics[width=8.5cm]{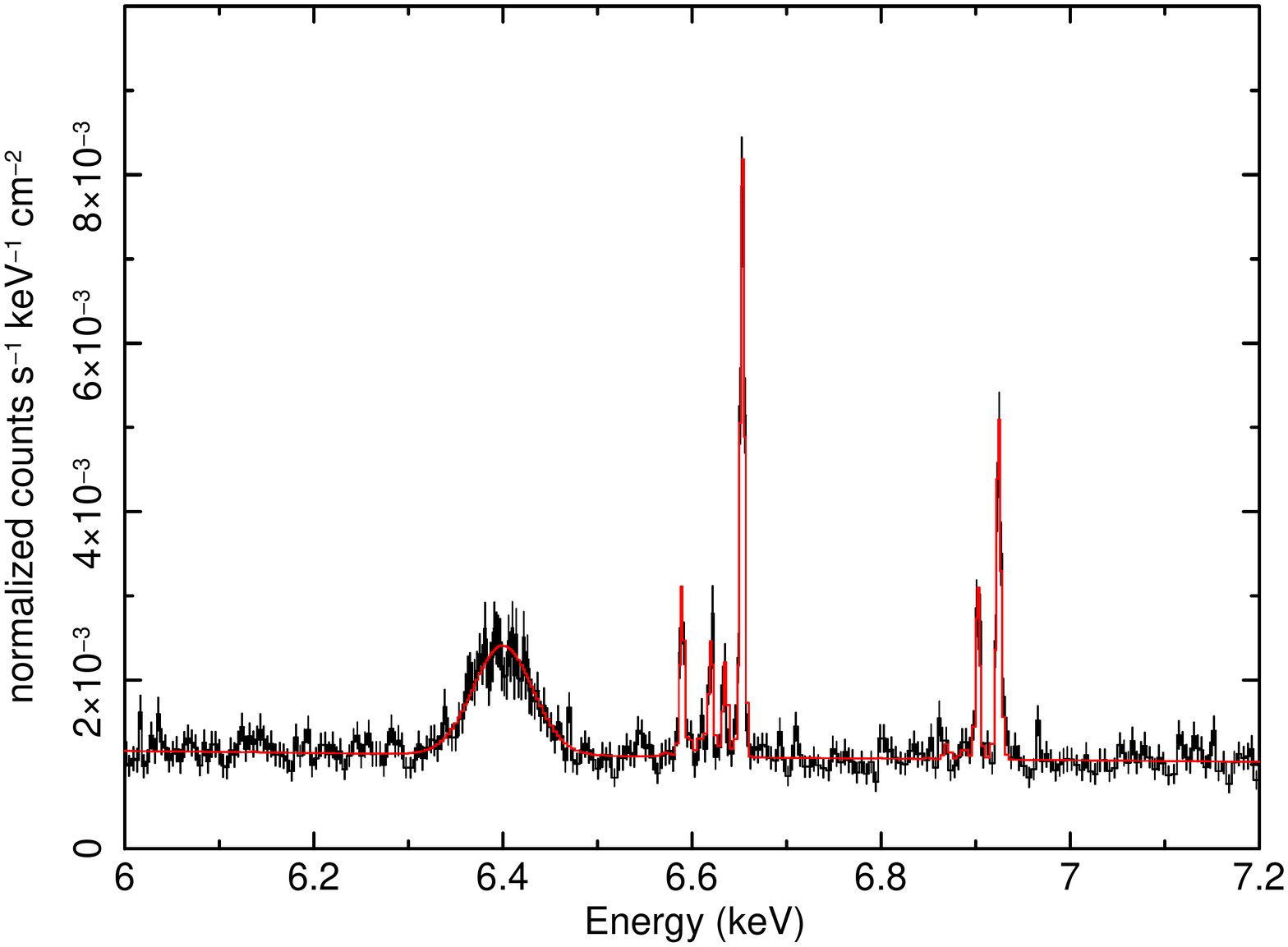}
\vspace{-5mm}
\caption{\footnotesize {\it Left}: A 150 ks {\it Chandra}/HETG first-order spectrum
  of the HMXB pulsar LMCX-4.  After strong binning, there is
  evidence of a neutral Fe K emission line at 6.4 keV, an He-like line
  at 6.7~keV, and potential for an H-like line at 6.97 keV.
  When fit with an ``apec'' model (shown in red), a modest red-shift
  of 900~km s$^{-1}$ is observed, perhaps indicating a specific and
  phase-dependent flow pattern within the binary. {\it Right}: A simulated
  50~ks Resolve spectrum of LMC X-4.  The He-like line
  complex is easily resolved into the four expected constituent lines,
  and the H-like line is revealed as a doublet. The redshift of the
  ionized lines is readily apparent, and the width of the neutral Fe K
  line may trace the radial extent of the inner disk.}
  \label{ch4fig2}
\vspace{-2mm}
\end{figure}

These systems also afford the opportunity to study the winds of
massive stars \citep{martinez-nunez17}.  In particular,
sharp flux dips in the light curves of high-mass X-ray binaries likely
signal obscuration by dense clumps that are characteristic of UV
line-driven winds from massive stars, giving us a unique chance to
directly probe these clumps.  Most current data only suggest that the
obscuration events are the result of "grey" absorption, without a
specific energy dependence.  Indeed, it is not clear if the dips are
due to a single cloud with a fairly uniform density, or a collection
of very dense clumps with a low volume filling factor.

The exquisite velocity resolution enabled by
{\it XRISM} will separate lines produced close to the compact object from
those excited in the larger stellar wind, and will detect velocity
shifts as a function of phase.  Resolve will also measure the
ionization, density and temperature of the accreting material,
enabling detailed comparisons to hydrodynamic simulations. It may be possible to detect lines and variability within the 
strongest flux dips, and to thereby determine the nature of the 
absorbers that cause the dips, improving our view of UV 
line-driven winds from massive stars.

\subsection{Ultra-luminous X-ray Sources}

Ultra-luminous X-ray sources (or, ULXs) can be defined as off-nuclear
point sources in nearby galaxies with implied luminosities above the
Eddington limit for a 10 solar mass black hole.  In recent years, a
growing number of ULXs appear to be strong pulsars, likely accreting
from massive stellar companions \citep{bachetti14,furst16,israel17a,israel17b}.  It is likely that the strong
magnetic fields play a role, and therefore it is not clear whether such sources are truly highly
super-Eddington, or less extremely super-Eddington (or
Eddington-limited) but subject to geometric and relativistic beaming.
Careful observations of mildly super-Eddington pulsars such as LMC X-4
may enable a better understanding of accretion in ULXs (see Figure~\ref{ch4fig2}).

In any super-Eddington source, radiation is coupled to gas and is -- by
definition -- defeating gravity, so very strong outflows are expected.
This is likely the case in the famous Galactic X-ray binary SS 433 \citep{kubota10,marshall13}.  In two ULXs, there is now tentative evidence
of fast, variable outflows in very deep observations with {\it
XMM-Newton} \citep{pinto16}. Resolve observations that are triggered on specific flux states or
behaviors in a subset of ULXs may be able to make wind detections and
to advance studies of the accretion flow in these enigmatic sources. 

With Resolve, we will begin to study whether
outflows have a single velocity, or a range of velocities, as well as
how and where they are launched. In a subset of the deepest ULX spectra obtained with {\it XMM-Newton}, there
is also tentative evidence of emission lines. It may be
possible with {\it XRISM} to examine whether these lines are excited in the wind of a
massive companion star, if they are tied to a young supernova remnant,
or if they are better attributed to local star formation.

\subsection{Magnetic Cataclysmic Variables}

When a neutron star or white dwarf has a magnetic field strong enough
to channel the accretion flow, it is directed towards the magnetic
poles.  Even with the power of present-day computers, simplifying
assumptions are necessarily involved in MHD simulations of accretion
flows in the presence of strong magnetic fields.  Details of a wide
range of phenomena --  from spin up/down of the compact object to ejection
of a wind (including in the propeller regime) --  depend on this
interaction.  Magnetic cataclysmic variables (CVs) of the intermediate polar
(IP) type usually have a partial accretion disk, and a well-developed
theory of X-ray emission \citep{hayashi14} that can
be used to measure a key parameter: the range of radii ($\Delta r$)
over which disk matter connects to the field lines. Since accretion
follows field lines that are close to dipolar, $\Delta r$ determines
the area on the white dwarf surface over which accretion occurs.

We anticipate that we can use the triplet ratio of the
He-like Fe lines to measure the density, expected to be in the
$10^{16-18}~ {\rm cm}^{-3}$ range \citep{mukai14}. Combined with
existing multiwavelength data, which allow us to infer other
parameters such as the inner disk radius and the total accretion rate,
the density measurement will allow us to infer $\Delta r$.

The He-like Fe lines of some magnetic CVs are extraordinary strong, and may indicate that resonant scattering plays a role \citep{terada04}. Since resonant scattering is
dependent upon geometry, we can expect to see the resonant component
and other components of the line change as a function of the white
dwarf spin phase.

%\newpage
\section{THE LIVES AND DEATHS OF STARS}
\label{stars}

{\it (Primary editors: Paul Plucinsky, L\'{i}a Corrales)}

\subsection{Star Formation and the Cold/Hot ISM}

%Molecular clouds are stellar nurseries, capable of bearing both low- and high-mass stars. These young stars actively interact with surrounding cloud materials, producing X-ray emitting hot plasmas at various spatial scales. {\it XRISM} is expected to detect individual emission lines from these plasmas, resolving the plasma flows accompanied by these interactions. This section introduces two science topics that highlight Resolve's unique capability; mass accretion onto newborn low-mass stars and diffuse X-ray plasma found in massive molecular clouds.

How molecular cloud materials accrete onto pre-stellar cores is a fundamental question in star formation. Optical, infrared, and radio observations now image accretion disks around young stellar objects \citep{alma15}, but they have not resolved the innermost regions where matter finally accretes onto the cores. Mass accretion at the innermost region is probably induced by interaction of large scale magnetic fields of the core and the innermost disk. Collimated jets seen in their outskirts are suspected to be byproducts of this innermost accretion process. 

%Young low-mass stars are known to emit strong X-rays via enhanced magnetic activities. Weak-line T Tauri stars in the late pre-main sequence stage drive a strong magnetic dynamo in the stellar interior with deep convection and differential rotation. Classical T Tauri stars in a younger stage also drive this activity, but they may also drive X-ray activities via shock heating of accretion matter onto their stellar surface --- 
%High-resolution spectroscopy of TW Hya with {\it Chandra's} gratings suggests the presence of relatively cool, high density plasma from highly ionized Ne and O emission lines at $\lesssim$1~keV \citep{kastner02,brickhouse10}.

%Although some mass accretion activity is seen in classical T Tauri stars, most stellar mass accretes during the earliest pre-main sequence stage when the pre-stellar cores are still heavily shrouded by thick CSM. Their X-ray emission is stronger than that of T Tauri stars, with hotter plasma temperatures and higher X-ray luminosities (e.g., {\it kT} $\sim$ 5~keV, {\it L$_{\rm X}$} $\sim$10$^{31}$~{\rm ergs~s$^{-1}$}, \cite{imanishi01}). Protostars drive violent magnetic activities, perhaps related to their strong mass accretion.

X-ray emission from protostars is heavily obscured by thick ({\it N$_{\rm H}$} $\geq$ 10$^{22}$~{\rm cm$^{-2}$}) circumstellar medium (CSM) below 2 keV where various emission lines are. However, their X-ray spectra often show strong Fe K emission lines at $\sim$6.7 keV as well as Fe fluorescence at 6.4 keV, which provide powerful tools to diagnose plasma conditions and dynamics of the innermost region. Freefall and Keplerian velocities near the core are a few hundred km s$^{-1}$, and hot plasmas are also expected to flow around these velocities. Unfortunately, existing high-resolution grating spectrometers are insufficient to resolve such velocity shifts. {\it XRISM} will resolve plasma motions near the core for the first time.

{\it XRISM} observations of protostars will reveal: i) geometry of X-ray plasmas around the core from Doppler shift or broadening measurements of highly ionized Fe emission lines, ii) plasma condition (e.g., equilibrium state) from multiple Fe emission lines, iii) dynamics of the innermost accretion disk from Doppler shifts or broadening of the fluorescent Fe K line, iv) jet formation mechanism from Doppler shifts of emission lines during X-ray flares.

Two nearby molecular clouds, $\rho$ Oph ($d \sim 165$ pc) and R CrA ($d \sim 130$pc), hold protostars that are bright enough in X-rays for detailed spectroscopy with {\it XRISM}. The other stellar classes important for mass accretion study are FU Ori/EX Lup type stars \citep{audard14}, which are young stars experiencing episodic mass accretion outbursts. The events occur rarely, but if such an outburst occurs in a nearby young star, Resolve should be able to collect valuable information on mass accretion activity.

\begin{figure}[htb]
\begin{center}
\includegraphics[width=1.0\textwidth]{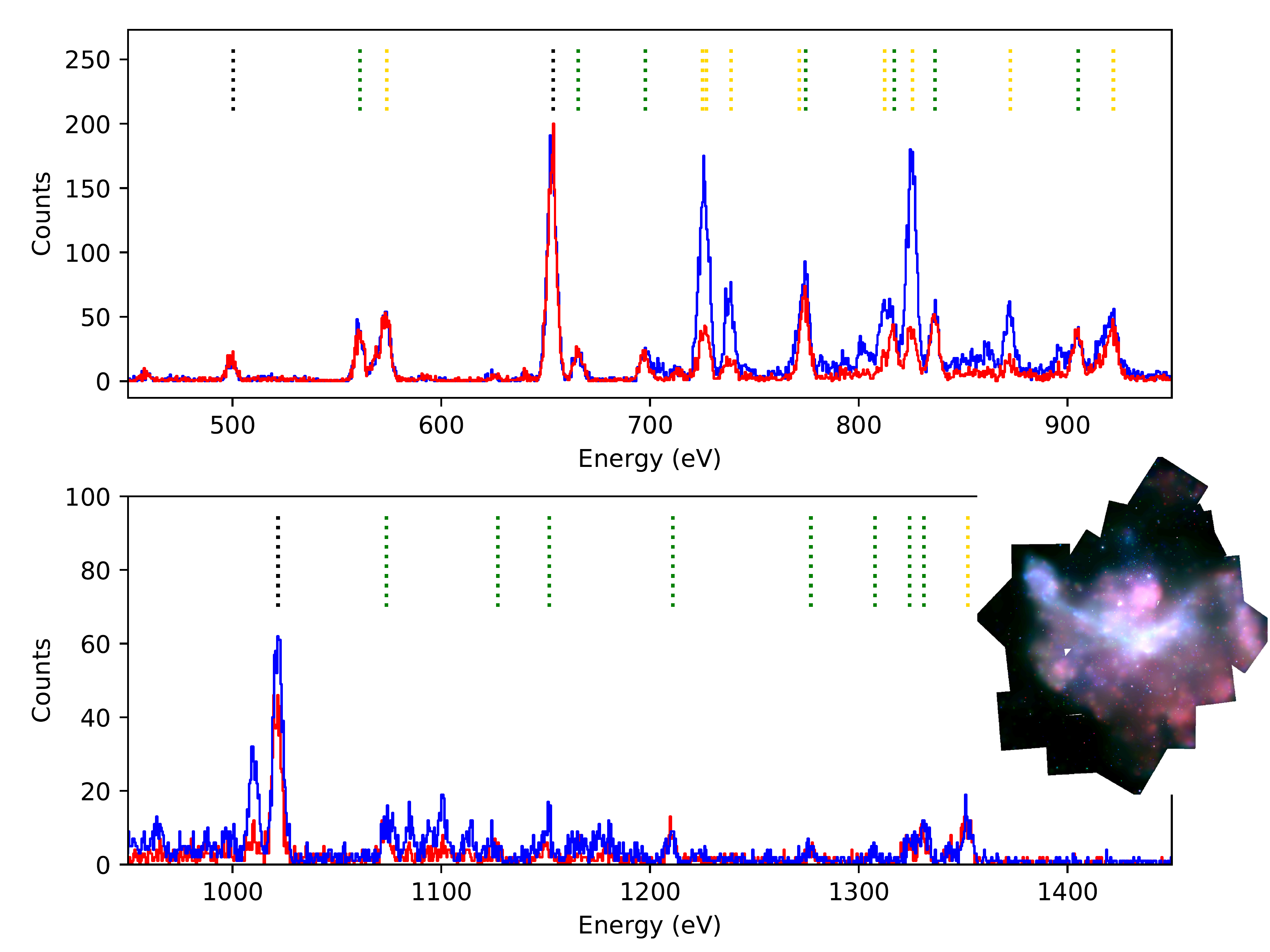}
\vspace{-4mm}
\caption{\footnotesize Simulated 30 ks Resolve spectra of two regions in the Carina nebula. The model assumes that half of the emission originates from hot, ionized plasma, and the other half from charge exchange of the same population of ions with cold gas. The green and yellow dotted lines indicate the positions of emission lines which are expected to originate primarily from charge exchange and thermal plasma, respectively; black dotted lines indicate emission lines expected to be have strong emission from both mechanisms. {\it Inset}: {\it Chandra} true color X-ray image of the region (Townsley et al., 2011), showing diffuse plasma extending over the Carina nebula as well as multiple X-ray bright young stars. Red: 0.50-0.70 keV, Green: 0.70-0.86 keV, Blue: 0.86-0.96 keV.}
\label{etacar}
\end{center}
\vspace{-6mm}
\end{figure}

Some massive star forming regions are accompanied by diffuse hot X-ray plasma with {\it kT} $\sim$0.1--0.8~keV (RCW38: \cite{wolk02}, M17: \cite{townsley03,hyodo08}, Carinae nebula: \cite{hamaguchi07,ezoe09,townsley11}, Orion nebula: \cite{gudel08}, see Figure~\ref{etacar}). The plasma apparently fills cavities of the parent molecular cloud and interacts with cold, thick molecular clouds at the contact surface. The origin of the diffuse plasma is not understood. One hypothesis is that the  plasma consists of multiple SNRs; however, none of them have a spherical shape like conventional SNRs or are accompanied by compact objects except for a few cases \citep{hamaguchi09,pires09}. The other hypothesis is that the diffuse plasma is thermalized by collision of winds from massive stars in the field. However, it is controversial if massive stars in the field can supply enough energy to thermalize the observed X-ray plasmas.

The keys to solving this problem should be in the elemental abundance distribution and the dynamics of the plasmas. X-ray CCD spectra of the Carina nebula demonstrated strong spatial variation of the Fe L emission lines \citep{hamaguchi07,ezoe09,townsley11}. If this is caused by Fe abundance variation, the X-ray plasmas are likely produced by supernova explosions. Meanwhile, supernova ejecta fly much faster than massive stellar winds. This difference may be seen in broadenings of emission line unless the plasmas are fully relaxed.

%The diffuse spectra show emission lines of various elements such as Fe, Si, Mg, Ne, O and N below $\sim$2 keV. XRISM/Resolve will resolve these emission lines individually for the first time, which are crucial for reliable abundance measurements. Resolve should be able to measure Doppler broadenings down to $\sim$1200~{\rm km~s$^{-1}$} with emission lines around 1 keV.

{\it XRISM} observations of the diffuse X-ray plasmas in massive star forming regions will reveal: i) plasma elemental abundances and their spatial variation by resolving individual emission lines, ii) plasma dynamics from Doppler shifts or broadening of these emission lines, iii) the presence of charge exchange lines\citep{townsley11}, which traces interaction between hot plasma and cold molecular gases. A discovery of turbulence in the X-ray plasma may also yield greater insight into the physics of particle acceleration in the star forming region.

\subsection{Supernovae and their Remnants}

%The physical insights that may be derived from X-ray spectroscopy of extended objects such as SNRs are limited by the moderate spectral resolution of the current generation of CCD-type detectors which are, for example, unable to resolve the lines that make up the He-like triplets.  One needs sensitive measurements of individual emission lines to fully characterize the state of the plasma. 

\begin{figure}[htb]
\begin{centering}
\includegraphics[width=1.0\textwidth]{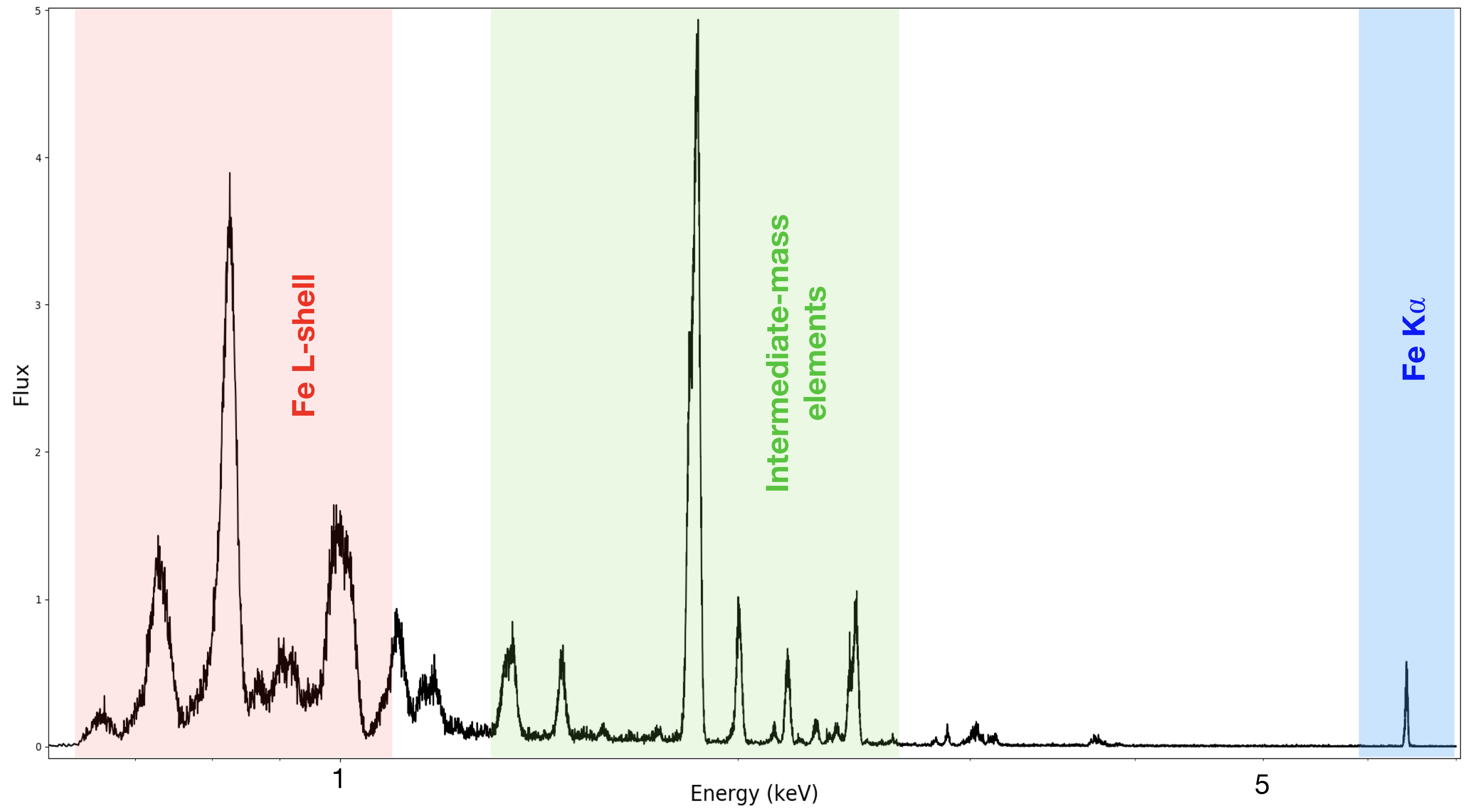}
\caption{\footnotesize A sample spectrum that one might expect from a young Type Ia SNR, such as Tycho or Kepler's SNR. Lines from various elements or groups are highlighted. Lines are broadened due to the extremely high temperatures that result from gas being shocked by a 5,000 km s$^{-1}$ shock wave. {\it XRISM} will resolve the widths of these lines in SNRs, leading to a direct measurement of the plasma temperature.}
\label{fig:tycho}
%\vspace{-4mm}
\end{centering}
\end{figure}

The high-resolution spectra that will be acquired by {\it XRISM} promise to revolutionize the spectral analysis of extended objects like SNRs, much as the spectra from the gratings instruments on {\it Chandra} and {\it XMM-Newton} have done for point-like sources. An example of the type of spectrum that {\it XRISM} can obtain from an SNR is shown in Figure~\ref{fig:tycho}, showing a simulated spectrum of a Type Ia remnant. The various line complexes are highlighted, and broadening of the lines by several thousand km s$^{-1}$ is included. We briefly describe here some of the science that will be enabled by this new capability. The reader is referred to the more comprehensive reviews written for {\it Astro-H} \citep{hughes14,long14,smith14} for a more complete list.

{\bf Type Ia SNRs} Type Ia SNe are generally believed to form through two scenarios: the single-degenerate (SD) channel  composed of a white dwarf (WD) and a stellar companion and the double-degenerate (DD) channel  composed of two WDs \citep{maoz14}.  Evidence has grown that both scenarios operate in nature but it is not clear if one channel is the dominant mechanism nor is it clear what the precise explosion mechanisms are for these systems.  

Theoretical work on the explosion mechanisms and the nucleosynthetic yields\citep{nomoto17,seitenzahl17} indicate that the amount of neutron-rich species ($^{58}$Ni, $^{56}$Fe, $^{54}$Fe, and $^{55}$Mn) produced is sensitive to the central density of the WD, the process referred to as neutron-rich nuclear statistical equilibrium (n-NSE).  The precise amount of these neutron-rich species and the intermediate mass elements (IME) also depends on the details of the explosion.  These different nucleosynthetic yields offer the possibility that X-ray spectroscopy of the remnant of a Type Ia SNe can distinguish a progenitor of a WD near the Chandrasekhar mass (SD scenario) from one with a lower central density (DD scenario). The Ni/Fe and Mn/Fe ratios in the {\it Suzaku} spectra of 3C 397 imply a SD progenitor, assuming that these ratios can only be achieved with high neutronization in the ejecta \cite{yamaguchi15}. The ratio of Ca to S was used as a diagnostic of high neutronization in the galactic remnants Tycho, Kepler, 3C 397 and G337.2-0.7 and the Large Magellanic Cloud remnant N103B to argue that the metallicity of the progenitor could not explain the high neutronization  \cite{martinez-rodriguez17}. %These results are based on CCD-resolution spectra of a handful of remnants with the associated large uncertainties from these moderate resolution spectra. For example, \cite{yamaguchi15} fit the Cr-K$\alpha$ , Mn-K$\alpha$ , Fe-K$\alpha$, and Ni-K$\alpha$ line complexes with a single Gaussian even though multiple lines with strengths that depend on the plasma conditions contribute to the complex because the CCD-resolution spectra do not allow the individual lines to be resolved.  

%Further complications are that the Mn-K$\alpha$ complex is partially blended with the Fe-K$\alpha$ complex, the Fe-K$\beta$ complex contributes to the Ni-K$\alpha$ complex, and the continuum level can not be well-determined in this line-rich region.   

The high-resolution spectra from Resolve of Type Ia SNRs will allow an accurate determination of the fluxes of all the lines that contribute in the Cr K$\alpha$, Mn K$\alpha$, Fe K$\alpha$, and Ni K$\alpha$ line complex bandpasses, allowing the ratios of these elements to be determined with unprecedented precision. In addition, Resolve spectra will cover the entire energy range from the IMEs to the Fe peak elements with the highest spectral resolution yet achieved.  These observations promise to be the definitive measurements of these abundances that will constrain the origin (SD vs. DD) and explosion mechanism for these objects. 

{\bf Core-Collapse SNRs} The explosion mechanism for CCSNe is still debated, with one common theme that spherical symmetry must be broken at some point to revive the shock which consistently stalls in simulations.  But where and how this asymmetry is produced is still unknown.  A key observational constraint needed is the 3D structure in space and velocity of the ejecta from the innermost layers of the star.  Pioneering work on the ejecta distribution of the brightest Galactic SNR Cas A has been done using {\it Chandra} data\citep{hughes00,delaney10}. Both these studies show a highly asymmetric ejecta distribution and significant overturning of the ejecta with high-Z material at larger radii than low-Z material. This distribution can be contrasted with that of Tycho, a Type Ia SNR, which shows a more symmetric distribution \citep{williams17,sato17}.  

The results to date have large uncertainties given the spectral resolution of the current generation of detectors.  For example, the studies with {\it Chandra} rely on the unresolved Si XII He-like triplet. Assumptions must be made about the temperature and ionization state to derive the centroid of the line complex.  With the resolution of {\it XRISM}, these assumptions will no longer be necessary. In fact, the resolution of the calorimeter is so high that for many lines of sight through the centers of remnants like Cas A and Tycho, it will be possible to separate redshifted material on the far side of the shell from blueshifted material on the near side of the shell.  

The {\it Hitomi} results from the LMC SNR N132D from just 3.7 ks of data offer a preview of what will be possible with {\it XRISM}. In this short observation \citep{hitomin132d}, the Fe K$\alpha$ complex was shown to be consistent with red-shifted emission with a velocity of $\sim 800$ km s$^{-1}$, while the S emission is consistent with the local ISM velocity in the LMC, indicating a highly asymmetric ejecta distribution. However, this result is model-dependent and significant at only the 1.6$\sigma$ level.  

\begin{figure}[htb]
\begin{center}
\includegraphics[width=1.0\textwidth]{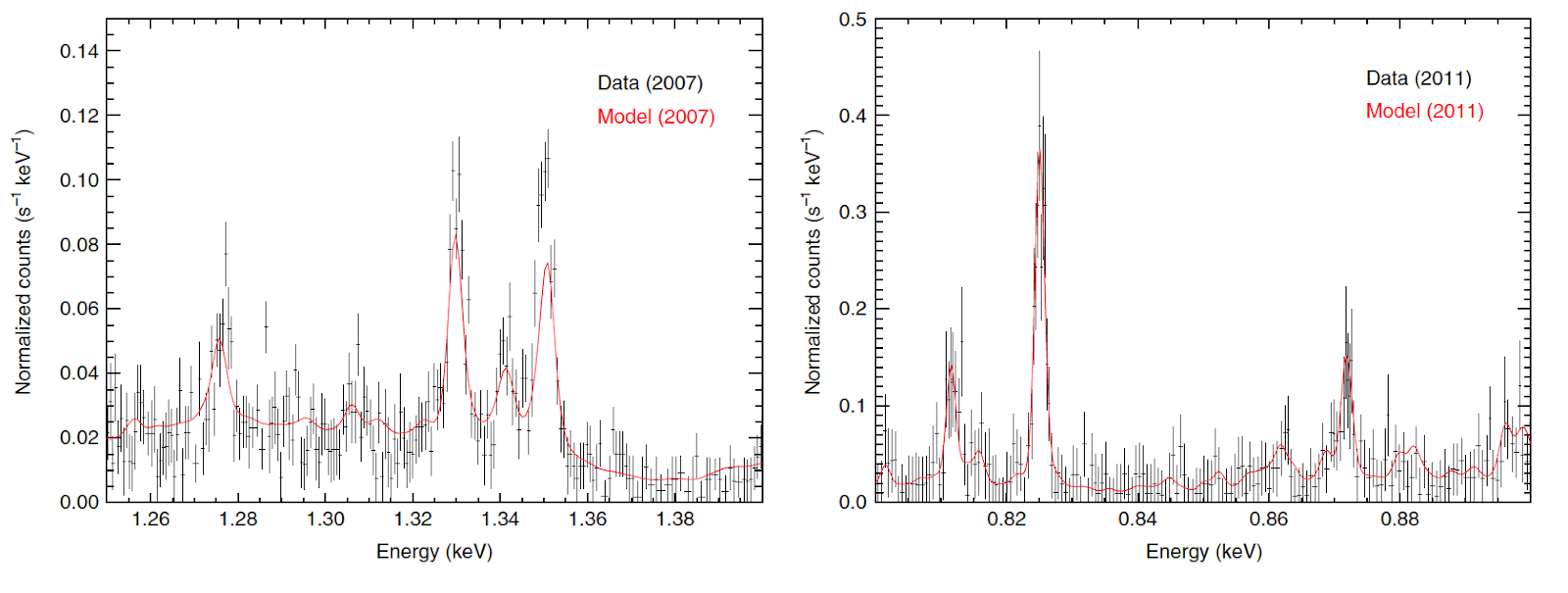}
\vspace{-3mm}
\caption{\footnotesize Fits to the Mg He-like triplet  (LEFT) and Fe-L region (RIGHT) in the {\it Chandra} HETG data from SN1987A yield the ion temperatures for these elements \citep{miceli19}.}
\vspace{-3mm}
\label{fig:pp1}
\end{center}
\end{figure}

%The nucleosynthetic yields for CCSNe are a function of the mass, metallicity, and rotation of the progenitor \citep{sukhbold16}. Several attempts have been made to measure the O, Ne, Mg, Si, and Fe abundances of young, ejecta dominated SNRs to constrain the mass of the progenitor. But all of these efforts have been hampered by a lack of precision due to the moderate spectral resolution of the instruments. The high-resolution spectra obtained by {\it XRISM} will usher in a new era where precise measurements will be used to constrain the properties of the progenitor and the type of compact object that forms.

\begin{wrapfigure}{L}{0.55\textwidth}
    \centering
    \vspace*{-2mm}
    \includegraphics[width=0.55\textwidth]{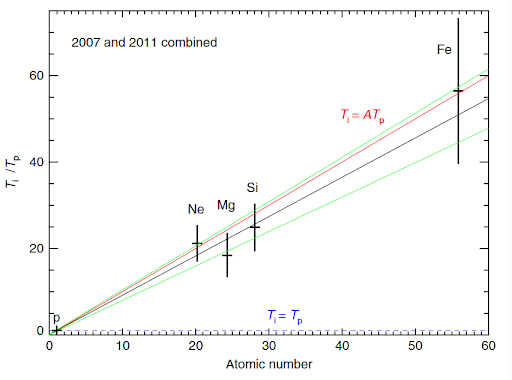}
    \caption{\footnotesize {Ion to proton temperature ratio as a function of atomic number in SN1987A (from Miceli et a. 2019).}}
    \vspace{-3mm}
\label{fig:pp2}
\end{wrapfigure}

{\bf Plasma Physics in SNRs} The processes which heat particles in shock waves is one of the most active areas of research in shock physics. It is not clear if the particles behind the shock will assume a temperature proportional to the mass of the particle, or if partial equilibration amongst the electrons, protons, and ions occurs (and if so, on what timescales). A recent result on SN 1987A \citep{miceli19} argues that the X-ray spectra are consistent with mass-proportional heating behind the shock front for the ions of Ne, Mg, Si, and Fe. This analysis compared the widths of the lines as determined by the HETG on {\it Chandra} to the predictions from 3D hydrodynamical models (see Figure~\ref{fig:pp1}). The ratio of the ion temperature to the proton temperature is shown as a function of Z in Figure~\ref{fig:pp2} (the analysis for Fe is based solely on the Fe L lines). {\it XRISM} will measure the widths of these lines to higher precision and allow more elements, such as Fe K$\alpha$, to be measured. The SN1987A analysis also assumes that the plasma conditions are similar for the entire remnant. For nearby galactic remnants, Resolve will measure the line broadening of ions as a function of position behind the shock, disentangling the effects of thermal broadening and bulk motions from mass-proportional heating.

\begin{figure}[h!tb]
\begin{center}
\includegraphics[width=1.0\textwidth]{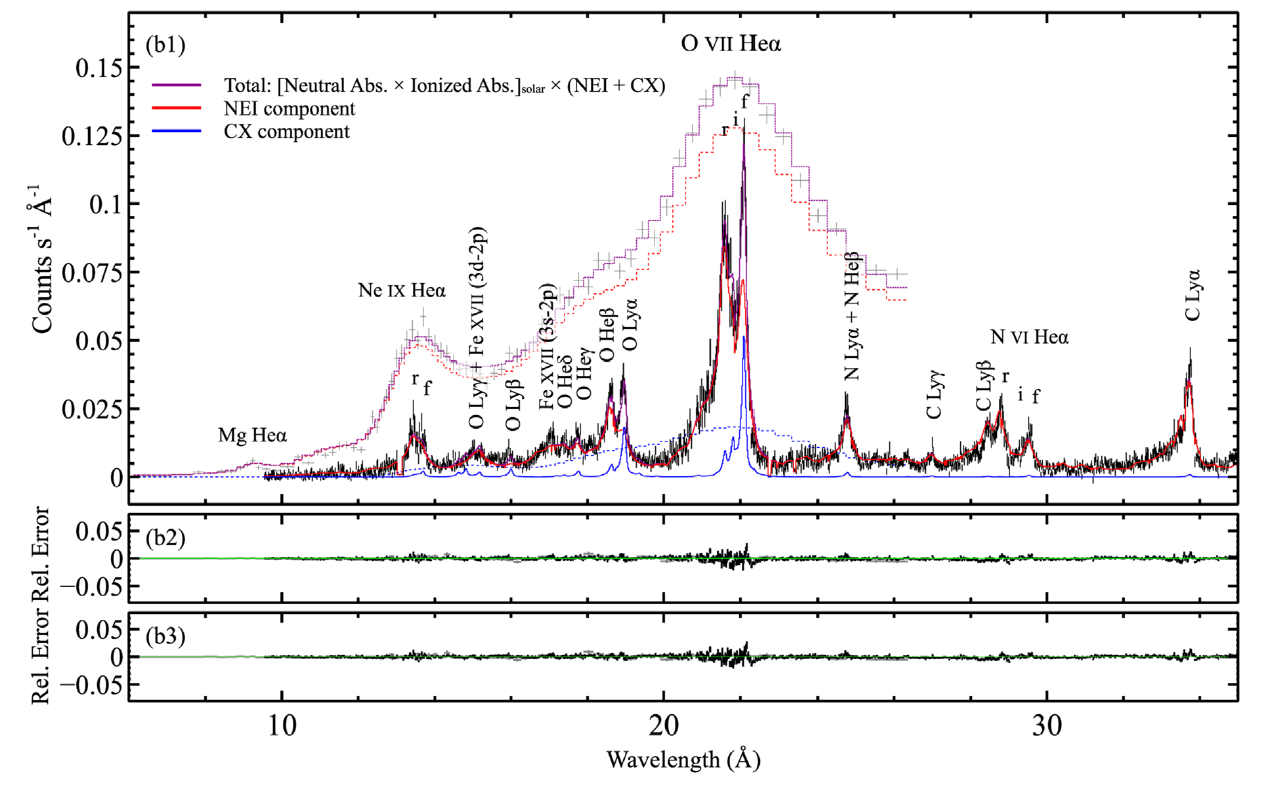}
\end{center}
\vspace{-5mm}
\caption{\footnotesize The {\it XMM-Newton} RGS spectrum of the SW-K knot in the Cygnus Loop. The black points are the data and the purple line is the best-fitted model including a significant charge exchange (CX) component and absorption from an ionized component.  The individual NEI (red) and CX (blue) components are shown. The upper data show the {\it XMM-Newton} pn CCD spectrum.  The residuals labelled as ?b2? are for a model in which the ionized absorber is along the line of sight while the residuals labelled as ?b3? are for an ionized absorber local to the SW-K knot itself.}
\vspace{-4mm}
\label{fig:pp3}
\end{figure}

Just as the importance and prevalence of radiative recombination emission was not appreciated until {\it Suzaku} provided moderate-resolution spectra with low background and large collecting area \cite{ozawa09,yamaguchi09}, spectroscopy of SNRs will change dramatically as new and unexpected features are discovered in high-resolution spectra. An example of this is CX emission in SNRs, expected to be significant when high-temperature ions interact with neutral material. The best example of CX emission in a SNR to date comes from the {\it XMM-Newton} RGS spectrum of a dense knot in the Cygnus Loop \citep{uchida19}.  Figure~\ref{fig:pp3} shows the high-resolution spectrum from the RGS and the moderate-resolution spectrum from the pn. This example demonstrates how an acceptable fit to the CCD spectra could be achieved while excluding the CX emission but such a model fails to fit the high-resolution data of the RGS. 

Another important effect which has been mostly ignored in spectral fitting of SNRs is resonance scattering, which can alter the ratio of the forbidden and resonance lines. If such an effect is ignored in the spectral fitting, the fitted parameters of the model will be systematically biased.  This is one possible explanation for the so-called ``low-abundance problem" in which the ISM abundances determined from fitting the forward shock emission in a SNR are systematically lower than those determined by other means.  The key to investigating this apparent discrepancy is to resolve the He-like triplets and develop a model which represents the high-resolution data. The Resolve spectra will transform the field by providing the data that demand that these processes are included in the spectral models, significantly improving our understanding of SNRs.

\subsection{Exoplanets}

It has become increasingly apparent $-$ through studies of our own solar system and hot Jupiters around other stars $-$ that magnetic and high-energy activity from planet host stars have a marked effect on planet atmospheres and habitability. X-ray observations of exoplanet host stars provide a marker of magnetic activity along with X-ray and UV flux, which affect exoplanet atmospheres through photoionization, photo-induced chemical reactions, and substantial mass loss \citep{lammer03,sanz-forcada10}. With the recent launch of {\it TESS}, we expect many Earth-sized planets to be discovered around sun-like stars out to 100 pc, or M dwarf stars out to 10 pc. This proximity means that exoplanet host stars are likely to be X-ray bright enough for high signal-to-noise observations with {\it XRISM}, opening the door for contributions to exoplanet science. 

{\it XRISM} could also provide a new avenue of research by observing planetary transits from X-ray bright exoplanet host stars. X-ray transits reveal the full extent of an exoplanet atmosphere and probe evaporation, because the X-ray absorption cross-sections of atmospheric gases are larger than the optical. Only one exoplanet transit has ever been observed in the X-ray band $-$ that of the hot Jupiter HD 189733b $-$ which showed a 6\% transit depth, a factor of three deeper than the optical transit \citep{poppenhaeger13}. Xtend could also probe the atmospheres of hot Jupiter planets by measuring the depths of the X-ray transits, while Resolve can be used to search for high-resolution X-ray absorption features from volatile elements like C, N, and O \citep{wolk19}. Such measurements would be valuable for understanding atmospheric evaporation and measuring the composition of hot Jupiter atmospheres, which can then be extended to the effects of high energy irradiation and coronal mass ejection on Earth-like planets.

\section{{\it XRISM} IN THE LANDSCAPE OF 2020s ASTRONOMY}
{\it (Primary editors: Erin Kara, Mar\'{i}a D\'{i}az Trigo)}\\

\begin{figure}[htb]
\includegraphics[width=1.0\textwidth]{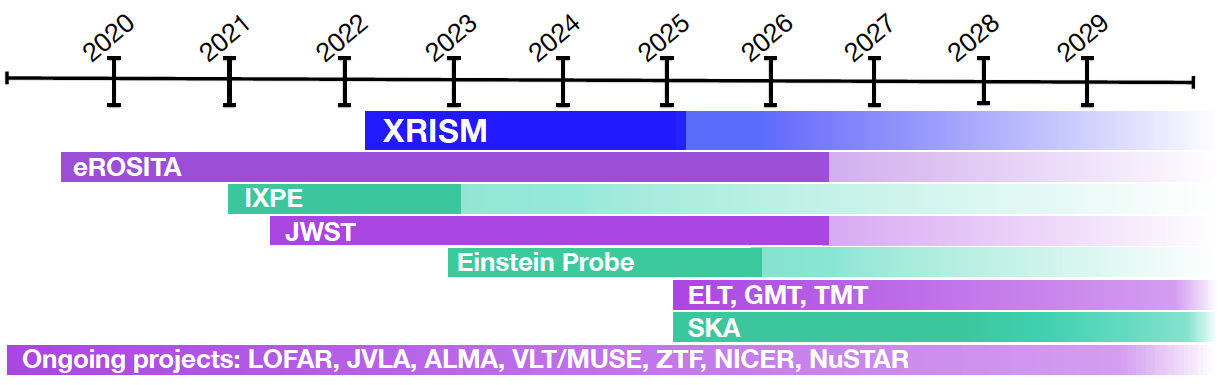}
\caption{\footnotesize A timeline of {\it XRISM} and current and future missions that are discussed in this chapter.}
  \label{futuremissions}
\end{figure}

With a suite of new observatories launching in the early 2020s, {\it XRISM} will benefit from a rich
landscape, and will add depth to many of the main science drivers of those future observatories (see Figure~\ref{futuremissions}). In
this chapter, we briefly highlight some of the science topics where {\it XRISM}, together with space- and ground-based
observatories, will make important breakthroughs, starting from the nano-parsec scales around black holes and
zooming out to mega-parsec scales in clusters of galaxies. We emphasize that the synergistic science outlined in
the following sub-sections is by no means exhaustive, and encourage further collaborative efforts.

\subsection{Independent Measurements of Black Hole Spin}

The strong gravitational field around a black hole can be completely parametrized by its mass and spin, which
imprint distinct signatures on the inner accretion flow that emits in the X-ray band. The spin tells how
black holes form, as a slowly rotating black hole can only reach the maximum spin allowed by general relativity
by increasing its mass \citep{bardeen70}. Therefore, measuring the distribution of black hole spin in supermassive
systems reveals the extent to which they are grown through accretion or mergers \citep{volonteri05}. On
the other hand, stellar mass black holes do not accrete enough material to affect their spin, and therefore,
measurements of black hole spin in these systems can constrain supernova mechanisms and the evolution of
massive stars.

Several techniques to measure black hole spin have been used. {\it XRISM} provides the best energy resolution in the Fe K$\alpha$ band, and thus will constrain the systematic uncertainties on the spin through broad line spectroscopy. {\it XRISM} will determine if other components, e.g., ionized winds, are imprinted on the broad Fe K$\alpha$ emission line. {\it IXPE} will provide the first X-ray polarization measurements of black hole spin. In black hole X-ray binaries in the thermal-dominated state, a fraction of the thermal flux will be deflected by the strong-field gravity of the black hole and will scatter off of the disk. In highly spinning black holes, the thermal emission will be highly polarized \citep{schnittman09}. Finally, observations with {\it NICER} can provide spin constraints through timing analysis, by measuring short timescale X-ray reverberation lags from reprocessed emission off the inner accretion disk \citep{kara19} or through high-frequency quasi-periodic oscillations in black hole X-ray binaries that may be associated with the orbital timescale at the innermost stable circular orbit \citep{motta14}. Together, {\it XRISM, IXPE} and {\it NICER} will use all of the information carried in photons (their energies, polarizations and arrival times) to put the tightest constraints on strong gravity in accreting black holes.

\subsection{The Disk-Wind-Jet Connection in Galactic Black Holes}

Accretion remains a central topic in modern astrophysics, and the variability timescales in X-ray binaries (ms to days/months) are particularly suited for studying the complex coupling between accretion and ejection. {\it XRISM} will measure the wind ionization, column density, and outflow velocity to derive the mass outflow rate and the associated mechanical power. Radio and IR facilities, such as the JVLA, ALMA and the VLT will measure the power output from jets. Comparing the power output from winds vs. jets, and their evolution during an accretion outburst \citep{fender16} will determine the relation between these different ejection modes, and will establish if they are important for triggering state transitions \citep{shields86}.

Observations of accreting black holes will test theoretical models that associate the X-ray corona
with the base of a relativistic jet \citep{markoff05} by making detailed measurements of the X-ray reflection features. {\it XRISM's} exquisite spectral resolution will be instrumental in disentangling these reflection features from other processes, such as absorption through disk winds. Hard X-ray observations with {\it NuSTAR} will provide an important ``lever-arm'' for {\it XRISM}, by putting simultaneous constraints on the X-ray continuum and reflection above 10 keV.

\subsection{Dust Scattering Halos and Molecular Clouds}

Studies of dust scattering halos around X-ray binaries constitute a useful tool to investigate the distribution and
properties of dust grains along the line of sight or to calculate distances via changes in the halo profile \citep{corrales15}. {\it XRISM} will allow a two-tiered approach to dust studies by using Xtend and Resolve
to investigate dust scattering haloes and the X-ray absorption fine structure at 0.2$-$2 keV, respectively. Adding
observations of CO with NOEMA or ALMA will help to constrain dust models by providing
an accurate map of the existent molecular clouds along the line of sight\citep{smith06,heinz15,kalemci18}.

\subsection{Following Extreme Transients in Time Domain Surveys}

X-ray observations have played a major role in time domain astronomy: from the
discovery of new galactic black holes, giant stellar flares, tidal disruption events (TDEs), and 
flares from Sgr A*, to the existence of super-Eddington accretors, the discovery of magnetars, and millisecond X-ray pulsars. Time domain astronomy is a quickly advancing field. Even in the next few years, we will see major increases in the rate of transient discovery. In the optical, ZTF and LSST will discover thousands of transients per night. In the radio, SKA will find tens of transients per night. In X-rays, the {\it eROSITA} all-sky survey will provide a 30-fold improvement in sensitivity, discovering faint transients, and producing a baseline against which we will compare future observations. Finally, the Chinese Academy of Science's Einstein Probe (to launch in late 2022) will use ``lobster-eye" optics to make a vast improvement in sensitivity for X-ray transient searches.

{\it XRISM} will follow-up on a wide variety of transients like TDEs, changing-look AGN, X-ray
bright SNe, Galactic black hole outbursts, GRB afterglows, and perhaps nearby kilonovae. {\it XRISM}
has a niche in time domain astronomy, and will be best suited for studying the detailed physics of bright, isolated
transients that evolve on timescales of days to weeks. High-resolution spectroscopy has already proved important
for transient science, with the discovery of highly ionized winds in the thermal state in black hole X-ray binaries
\citep{ponti12}, along with the discovery of disk winds (even ultrafast outflows) in TDEs \citep{miller15b}. {\it XRISM} will have a ToO response time similar to other large observatories like {\it XMM-Newton} and {\it Chandra}.

\subsection{Galactic Outflows and the Startburst-AGN Connection}

The discovery of massive dusty and molecular components in multi-phase galactic outflows associated with starbursts and AGN over the past decade confirmed that they have the potential to regulate and perhaps halt star formation in their host galaxies, but we still do not know which types of outflows most effectively disrupt and remove star-forming gas \citep{bae17}. Moreover, outflows can compress gas and trigger star formation, and it remains unclear whether starbursts trigger quasars or vice versa. The source of outflows in systems with both a starburst and an AGN is also an open question. Sensitive sub-mm facilities, such as ALMA and IRAM, as well as optical and near-IR integral field spectrographs, have illustrated the diversity of outflows \citep{rupke18}. The largest missing piece is measurements from hot and/or highly ionized gas, which contains most of the energy and much of the mass and metals \citep{cicone18}.

X-ray observations of outflowing gas on both small ($<$1 pc) and large ($>$10 kpc) scales will determine the
genesis and ultimate impact of outflows. In particular, Fe XXV (6.7 keV) and Fe XXVI (6.96 keV) trace both
the very hot gas driving starburst winds \citep{strickland09,mitsuishi11,liu14}
and the relativistic disk winds driven by the AGN \citep{tombesi13,tombesi15}. {\it XRISM} will provide
the first velocity, temperature, and mass measurements in multiple hot starburst winds, and will be sensitive to
weak and variable ultra-fast outflows in many AGNs. On larger scales, soft X-ray lines trace the expanding winds and their interaction with the ISM and CGM. %While {\it XRISM} has worse spectral resolution than existing grating spectrographs below 1 keV, it will provide the first spatially resolved velocity and plasma diagnostic maps for nearby winds, such as M82 or NGC 6240. The ionization balance and plasma state will also determine whether a given wind (or component) is powered by a starburst or AGN.

These measurements are important on their own, but must be understood in
the context of observations of cooler gas. The combination of {\it XRISM, JWST}, ALMA, and ground-based optical
and near-IR instruments (such as VLT/MUSE, Gemini/GMOS, TMT/IRIS, or ELT/HARMONI) will provide a detailed accounting of the physical state, bulk velocity, and mass of the neutral and ionized gas in
nearby (d $<$200 Mpc) starburst and AGN outflows, enabling us to determine how much gas can actually escape
the galaxy's gravity. In addition to the hottest gas, this will include cool H$_{2}$ detected in CO bands, warm H$_{2}$ detected in the 2.124 $\mu$m line, dusty outflows detected via PAH bands from 3$-$20 $\mu$m, warm ionized gas via Br$\gamma$, and hot, but cooling, gas from mid-IR forbidden lines, such as [O IV], [Ne V], [Ne VI], and [S IV].

\subsection{AGN Feedback in Clusters of Galaxies}

The hot atmospheres of many galaxy clusters are so luminous in X-ray emission that they should radiate away
their thermal energy in $<$ 1 Gyr. Uninhibited cooling should then fuel molecular gas and star formation at
rates of hundreds to thousands of solar masses per year. However, observed molecular gas reservoirs and star
formation rates are ten times lower than predicted from pure cooling. Instead, jets launched from the central
SMBH are thought to regulate cooling \citep{mcnamara07,mcnamara12}, but the details of
the AGN feedback process are not well understood. {\it XRISM} will allow us to probe the physics of AGN feedback
by measuring the velocity of bulk and turbulent motions of the hot phase of the ICM.

Molecular gas is another key component in the AGN feedback cycle. This is the reservoir for gas condensing
out of the hot atmosphere, and the fuel source for accretion onto the central supermassive black hole. Together {\it XRISM} and ALMA will probe the dynamics and energetics of the hot phase and
cold phases of the ICM, which is crucial for developing a complete model of AGN feedback.

\subsection{Particle Acceleration in Large-Scale Plasmas}

{\it XRISM} will provide the first direct measurements of the turbulent
line broadening in clusters that host extended radio haloes. These will be crucial to verify existing, indirect
hints that link the non-thermal component and the kinetic energy in the ICM \citep{eckert17,bonafede18}, opening a new window into quantitative studies of turbulent particle acceleration and magnetic field
amplification on large scales. Progress in this field is currently driven at a rapid pace by sensitive radio surveys
using LOFAR \citep{shimwell19} and GMRT \citep{intema17}, and will continue in the future with the
upgraded LOFAR 2.0 and uGMRT, as well as MWA, MeerKAT, and the SKA. As these surveys reveal more
and more of the faint non-thermal plasma and map its radio spectral index, the Resolve instrument will
deliver the complementary imaging spectroscopy capabilities required to connect the properties of the relativistic
electrons seen in the low-frequency radio band with detailed dynamics of the thermal gas permeating the cosmic
web, thus revealing the intricate physics behind the Universe's largest particle accelerators.

\subsection{Conclusion}

The importance of high-resolution X-ray spectroscopy to the future of astronomy in general cannot be overstated. Despite the fact that {\it Hitomi} only operated for a matter of weeks in 2016, the results from that spacecraft are among the most highly-cited papers in the field of high-energy astronomy in the past several years. By recovering the science lost by the untimely demise of {\it Hitomi}, {\it XRISM} will continue its legacy into the next decade, paving the way for future X-ray missions, such as {\it Athena}, which will be equipped with even greater capabilities.

The checkout, calibration, and performance verification phases will last approximately nine months, after which {\it XRISM} will enter the general observer phase. Astronomers around the world will be able to propose to an annual announcement of opportunity for competitively-awarded time on the observatory. The science that is briefly outlined in this white paper is only the beginning of what will be possible given the capabilities that this observatory will provide.

\newpage
\section{Acknowledgements}

Additional inputs to this white paper were provided by Hideki Uchiyama, Masayoshi Nobukawa, Kumiko K. Nobukawa, Shinya
Nakashima, Shigeo Yamauchi, Takeshi G. Tsuru, Edmund Hodges-Kluck, Iurii Babyk, Kenji Hamaguchi, Maurice Leutenegger, Renata Cumbee, and Aurora Simionescu. The document was edited and compiled by Brian Williams. \\

\noindent
The {\it XRISM} Science Team consists of the following members:\\

\noindent
Lorella Angelini (NASA/GSFC), Marc Audard (University of Geneva), Hisamitsu Awaki (Ehime University), Aya Bamba (University of Tokyo), Ehud Behar (Technion \& ISAS/JAXA), Laura Brenneman (Harvard-Smithsonian Center for Astrophysics), Greg Brown (Lawrence Livermore National Laboratory), L\'{i}a Corrales (University of Michigan), Elisa Costantini (SRON), Renata Cumbee (NASA/GSFC), Mar\'{i}a D\'{i}az-Trigo (ESO), Chris Done (University of Durham), Tadayasu Dotani (ISAS/JAXA), Ken Ebisawa (ISAS/JAXA), Megan Eckart (Lawrence Livermore National Laboratory), Dominique Eckert (University of Geneva), Satoshi Eguchi (Fukuoka University), Teruaki Enoto (Kyoto University), Yuichiro Ezoe (Tokyo Metropolitan University), Ryuichi Fujimoto (Kanazawa University), Yutaka Fujita (Osaka University), Yasushi Fukazawa (Hiroshima University), Akihiro Furuzawa (Fujita Health University), Luigi Gallo (Saint Mary's University), Liyi Gu (RIKEN), Matteo Guainazzi (ESTEC), Kouichi Hagino (Tokyo University of Science), Kenji Hamaguchi (NASA/GSFC), Isamu Hatsukade (University of Miyazaki), Takayuki Hayashi (NASA/GSFC), Kiyoshi Hayashida (Osaka University), Natalie Hell (Lawrence Livermore National Laboratory), Junko Hiraga (Kwansei Gakuin University), Edmund Hodges-Kluck (NASA/GSFC), Ann Hornschemeier (NASA/GSFC), Akio Hoshino (NASA/GSFC), Yuto Ichinohe (Rikkyo University), Manabu Ishida (ISAS/JAXA), Kumi Ishikawa (ISAS/JAXA), Yoshitaka Ishisaki (Tokyo Metropolitan University), Jelle Kaastra (SRON), Timothy Kallman (NASA/GSFC), Erin Kara (MIT), Satoru Katsuda (Saitama University), Richard Kelley (NASA/GSFC), Caroline Kilbourne (NASA/GSFC), Takao Kitaguti (RIKEN), Shunji Kitamoto (Rikkyo University), Shogo Kobayashi (Tokyo University of Science), Takayoshi Kohmura (Tokyo University of Science), Aya Kubota (Shibaura Institute), Maurice Leutenegger (NASA/GSFC), Mike Loewenstein (NASA/GSFC), Yoshitomo Maeda (ISAS/JAXA), Maxim Markevitch (NASA/GSFC), Hironori Matsumoto (Osaka University), Kyoko Matsushita (Tokyo University of Science), Dan McCammon (University of Wisconsin), Brian McNamara (University of Waterloo), Eric Miller (MIT), Jon Miller (University of Michigan), Ikuyuki Mitsuishi (Nagoya University), Tsunefumi Mizuno (Hiroshima University), Koji Mori (University of Miyazaki), Hideyuki Mori (NASA/GSFC), Koji Mukai (NASA/GSFC), Hiroshi Murakami (Tohoku Gakuin University), Richard Mushotzky (University of Maryland), Hiroshi Nakajima (Kanto Gakuin University), Kazuhiro Nakazawa (Nagoya University), Kumiko Nobukawa (Nara Women's University), Masayoshi Nobukawa (Nara University of Education), Hirofumi Noda (Osaka University), Hirokazu Odaka (University of Tokyo), Takaya Ohashi (Tokyo Metropolitan University), Masahiro Ohno (Hiroshima University), Takashi Okajima (NASA/GSFC), Naomi Ota (Nara Women's University), Stephane Paltani (University of Geneva), Rob Petre (NASA/GSFC), Paul Plucinsky (Harvard-Smithsonian Center for Astrophysics), F. Scott Porter (NASA/GSFC), Katja Pottschmidt (NASA/GSFC), Kosuke Sato (Saitama University), Toshiki Sato (RIKEN/NASA GSFC), Makoto Sawada (RIKEN), Hiromi Seta (Tokyo Metropolitan University), Megumi Shidatsu (Ehime University), Aurora Simionescu (SRON), Randall Smith (Harvard-Smithsonian Center for Astrophysics), Yang Soong (NASA/GSFC), Yasuharu Sugawara (ISAS/JAXA), Andy Szymkowiak (Yale University), Hiromitsu Takahashi (Hiroshima University), Toru Tamagawa (RIKEN), Takaaki Tanaka (Kyoto University), Makoto Tashiro (Saitama University), Yukikatsu Terada (Saitama University), Yuichi Terashima (Ehime University), Yohko Tsuboi (Chuo University), Masahiro Tsujimoto (ISAS/JAXA), Hiroshi Tsunemi (Osaka University), Takeshi Tsuru (Kyoto University), Hiroyuki Uchida (Kyoto University), Hideki Uchiyama (Shizuoka University), Yoshihiro Ueda (Kyoto University), Yuusuke Uhida (Hiroshima University), Shinichiro Uno (Nihon Fukushi University), Jacco Vink (University of Amsterdam), Shin Watanabe (ISAS/JAXA), Brian Williams (NASA/GSFC), Shinya Yamada (Tokyo Metropolitan University), Hiroya Yamaguchi (ISAS/JAXA), Kazutaka Yamaoka (Nagoya University), Noriko Yamasaki (ISAS/JAXA), Makoto Yamauchi (University of Miyazaki), Shigeo Yamauchi (Nara Women's University), Tahir Yaqoob (NASA/GSFC), Irina Zhuravleva (University of Chicago)\\

\newpage

\bibliography{refs}
\bibliographystyle{naturemag}
\nocite{*}

\end{document}